\documentclass[journal=jacsat,manuscript=article]{achemso}

\usepackage[version=3]{mhchem} 
\usepackage{amsmath}
\usepackage{braket}
\usepackage{xcolor}
\usepackage{amsthm,amsmath,amssymb}
\usepackage{physics}
\usepackage{bm}

\usepackage{accents}
\newlength{\dhatheight}
\newcommand{\doublehat}[1]{%
    \settoheight{\dhatheight}{\ensuremath{\hat{#1}}}%
    \addtolength{\dhatheight}{-0.35ex}%
    \hat{\vphantom{\rule{1pt}{\dhatheight}}%
    \smash{\hat{#1}}}}

\author{Yu Wang}
\affiliation{Department of Chemistry, School of Science, Westlake University, Hangzhou 310024 Zhejiang, China}
\altaffiliation
{Institute of Natural Sciences, Westlake Institute for Advanced Study, Hangzhou 310024 Zhejiang, China}

\author{Ruihao Bi}
\affiliation{Department of Chemistry, School of Science, Westlake University, Hangzhou 310024 Zhejiang, China}
\altaffiliation
{Institute of Natural Sciences, Westlake Institute for Advanced Study, Hangzhou 310024 Zhejiang, China}

\author{Wenjie Dou}
\email{douwenjie@westlake.edu.cn}
\affiliation{Department of Chemistry, School of Science, Westlake University, Hangzhou 310024 Zhejiang, China}
\altaffiliation
{Institute of Natural Sciences, Westlake Institute for Advanced Study, Hangzhou 310024 Zhejiang, China}

\title
  {Manipulating nonadiabatic dynamics by plasmonic nanocavity}

\begin{document}








\begin{abstract}
In recent years, plasmonic nanocavities have emerged as powerful tools for controlling and enhancing light-matter interactions at the nanoscale. This study explores the role of plasmonic nanocavities in manipulating nonadiabatic dynamics, particularly in systems where fast electronic transitions are crucial. By coupling molecular states to the plasmonic resonances of metallic nanocavities, we demonstrate that the local electromagnetic fields generated by plasmons can significantly influence the rates and pathways of nonadiabatic transitions, including electron transfer and excitation relaxation processes. Using the Floquet quantum master equation (FQME) and Floquet surface hopping (FSH) methods that we previously developed, we find that plasmonic nanocavities can enhance nonadiabatic effects by tuning the plasmonic coupling strength, the molecule-metal interaction strength, and the material properties.
These approaches offer a new perspective for predicting molecular dynamics in ultrafast processes. Our findings pave the way for designing novel plasmonic devices capable of controlling electron and energy transfer in chemical reactions, optoelectronic applications, and quantum information processing.
\end{abstract}

\section{1. INTRODUCTION}

A plasmonic nanocavity is a nanoscale structure that supports surface plasmon polaritons (SPPs)—oscillations of free electrons in a metal coupled with electromagnetic fields. These cavities can confine light to subwavelength scales, enhancing local electromagnetic fields\cite{kwon2014design,hugall2018plasmonic,bitton2022plasmonic,mandal2023theoretical}. Plasmonic nanocavities, such as metallic nanoparticles or sharp tips in gap structures, exhibit field enhancement at metal-dielectric interfaces or in nanoscale gaps\cite{ditlbacher2005silver,kress2015wedge,perney2007tuning,novotny2011antennas}.
Plasmonic volumes are typically much smaller than optical cavities, which means that the confined space within plasmonic cavities enables strong coupling with individual molecules. This leads to a regime where the interaction between the plasmonic field and a single molecule becomes the dominant factor.\cite{maccaferri2021recent}.
The plasmonic nanocavity mode is a standing wave where the intensity is confined, in contrast to the SPP mode, which is not localized and propagates along the interface, with its intensity exponentially decaying in the direction perpendicular to the interface.
It has been demonstrated that the plasmonic cavity mode exhibits superior activity compared to the SPP mode.\cite{lyu2023periodic,lyu2023plasmonic}.

At the heart of plasmonic nanocavities is the enhancement of electromagnetic fields, often resulting in localized ``hot spots" where the field intensity is significantly higher than that of the incident light. 
This intense field enhancement facilitates both sensing and light-matter interactions. For example, Raman scattering from molecules near the cavity can be significantly amplified, enabling ultrasensitive detection of molecular species. As a result, Surface-Enhanced Raman Spectroscopy (SERS) becomes a powerful tool for biosensing and chemical detection\cite{im2013self,maier2006plasmonic,li2017fabrication,sharma2014hybrid}.
In the field of chemical sensing, the localized field enhancement can accelerate reactions by lowering activation energies, enabling real-time monitoring of chemical processes with high temporal resolution\cite{ameling2010cavity,li2020real,kwon2018tunable}. This is particularly useful in catalysis, where plasmonic nanocavities can enhance the efficiency of chemical reactions, such as those involved in energy conversion or environmental remediation\cite{fojt2024controlling}.
Moreover, intense field enhancements are being explored in the realm of quantum optics and nonlinear optics\cite{marquier2017revisiting,fryett2017cavity,panoiu2018nonlinear}. Plasmonic nanocavities can facilitate enhanced nonlinear interactions at the nanoscale, opening avenues for the development of novel light sources, quantum communication systems, and high-performance optical sensors.

Although plasmonic nanocavities have been extensively studied in the context of light emission\cite{russell2012large}, catalysis\cite{lyu2023plasmonic}, and sensing\cite{butt2024review}, their potential to influence the electronic states and dynamics of molecules—especially during nonadiabatic transitions—remains not fully understood.
There have been several theoretical studies on photochemical and photophysical processes in plasmonic cavities. For example, Climent and co-workers\cite{climent2019plasmonic} show that plasmonic nanocavities can alter chemical reaction pathways in the plasmonic nanocavities
by means of electronic structure calculations. 
Fregoni and co-workers\cite{fregoni2021strong} combined a high-level quantum chemical description of molecules with a quantized model of localized surface plasmons to demonstrate how plasmonic nanocavities can manipulate molecular charge density, providing valuable benchmarks to guide the development of molecular polaritonics.
Mondal and co-workers\cite{mondal2022strong} present a method to estimate radiation-matter coupling, mode volume, mode lifetime, and quality factor for plasmonic cavities of various shapes.
Jamshidi and co-workers\cite{jamshidi2023coupling} simulate the complex nature of plasmonic nanocavities and their coupling with emitters within the framework of quantum dynamics, emphasizing the significant role that the plasmonic structure plays in the short-time dynamics of the molecules.
Theoretical calculations for plasmonic cavities are still relatively scarce because incorporating plasmonic cavities into theoretical studies poses significant challenges.
These systems involve light interacting with electronic states, leading to strong light-matter coupling and hybrid states like polaritons. Electron-phonon couplings further contribute to energy dissipation and decoherence. A robust theoretical framework must capture the complex and quantum coupling between these entities.
Furthermore, nonadiabatic effects are crucial for accurately describing processes such as energy transfer, relaxation, and chemical reactions within plasmonic cavities\cite{fregoni2022theoretical}.

In this study, we utilize the Floquet quantum master equation (FQME) and Floquet surface hopping (FSH) methods, developed in previous work\cite{wang2023nonadiabatic,wang2024nonadiabatic}, 
to investigate nonadiabatic relaxation processes in plasmonic nanocavities. Pyrazine is selected as the molecular system coupled to the cavity due to its widespread use as a model for nonadiabatic dynamics.
This is attributed to its closely spaced $S_1$ ($n\xrightarrow{}\pi*$) and  $S_2$ ($\pi\xrightarrow{}\pi*$) states, which exhibit strong coupling via conical intersections (CIs)\cite{seidner1992b,woywod1994characterization}.
We find that the plasmonic cavity effectively facilitates both pumping and relaxation processes. While electron transfer between the molecule and the metal surface influences state populations, it does not significantly impact the overall rates of pumping and relaxation dynamics. This is because we consider a weak coupling between the molecule and the metal surface.
The results obtained from FQME and FSH are  consistent with each other. 
Due to the high computational cost of FQME, the FSH method offers a more computationally efficient alternative.

\section{2. THEORY}
\subsection{MODEL HAMILTONIAN}
The general model Hamiltonian in our study consists of three components: the system part, which incorporates a time-periodic driving term; the bath part; and the system-bath coupling part:
\begin{equation}
    \hat{H}(t) = \hat{H}_S(t) + \hat{H}_B + \hat{H}_C
\end{equation}
\begin{equation}\label{Hs}
    \hat{H}_S(t) = \hat{H}_{el}
    (\hat{\boldsymbol{R}},t) + U_0(\hat{\boldsymbol{R}}) + \sum_{\alpha}\frac{\hat{\boldsymbol{P}}_{\alpha}^2}{2m_{\alpha}}
\end{equation}
\begin{equation}\label{Hel}
    \hat{H}_{el}(\hat{\bm{R}},t) = \sum_{i, j=0}^{2} \mathcal{H}_{ij}(\hat{\bm{R}}) \ketbra{\text{S}_i}{\text{S}_j} + \left(\bm{\mu} \cdot \bm{E}(t) \ketbra{S_0}{S_2} + \text{h.c.}\right) + E_d(\hat{\bm{R}}) \hat{d}^{\dagger} \hat{d}
\end{equation}
\begin{equation}
    \hat{H}_B = \sum_{k}\epsilon_k \hat{c}_k^+ \hat{c}_k
\end{equation}
\begin{equation}
    \hat{H}_C = \sum_{k}V_{k}(\hat{c}_k^+ \hat{d} + \hat{d}^+ \hat{c}_k)
\end{equation}
Here, $\hat{H}_S$ represents the nuclear and electronic degrees of freedom (DOF) of pyrazine under classical light-matter interaction.
Specifically, $\mathcal{H}_{ij}$ accounts for the interaction among the ground state ($S_0$) and excited states ($S_1$ and $S_2$) of pyrazine.
For pyrazine, the transition dipole moment between $S_0$ and $S_2$ states is much larger than that between $S_0$ and $S_1$ states. Therefore, we focus exclusively on the transition dipole moment between $S_0$ and $S_2$ states, denoted by $\boldsymbol{\mu}$, which interacts with an electromagnetic field of strength $E$ and frequency $\Omega$, expressed as $\bm{E}(t)=\boldsymbol{E}\cos(\Omega t)$.
Here, the term $\boldsymbol{\mu}\cdot \boldsymbol{E}\cos({\Omega t})$ can be expressed as $A\cos({\Omega t})$, representing a time-periodic driving force with an overall strength of $A$.
Therefore, $\hat{H}_S(t)=\hat{H}_S(t+T)$ with the driving period $T$.
Additionally, a radical anion state $S_0^-$ is introduced by adding an electron to the lowest unoccupied molecular orbital (LUMO) of the $S_0$ state.
This is represented using second quantization notation, where $\hat{d}^+$ ($\hat{d}$) denote the creation (annihilation) operators for the effective spin-up electronic orbital in the LUMO (as shown in the electronic configuration in Figure S1).
The term ``effective spin-up electronic orbital" is used because we assume that the energy of this orbital ($E_d$) is equal to the energy difference between the radical anion state and the ground state of pyrazine. Simply considering the addition of an electron to the LUMO is insufficient to achieve the stable pyrazine anion.
Under the consideration of the $S_0^-$ state, the $\hat{H}_{el}$ has the matrix form:
\begin{equation}\label{1e-Ham}
\hat{H}_{el}(\hat{\boldsymbol{R}},t) = 
\begin{pmatrix}
h_0^- & 0 & 0 & 0 \\
0 & h_0 & 0 & \boldsymbol{\mu}\cdot \boldsymbol{E}(t)\\
0 & 0 & h_1 & \lambda Q_c\\
0 & \boldsymbol{\mu}\cdot \boldsymbol{E}(t) & \lambda Q_c & h_2
\end{pmatrix}
\end{equation}
where $h_0^-$, $h_0$, $h_1$ and $h_2$ are relative energy of $S_0^-$, $S_0$, $S_1$, and $S_2$ states of the pyrazine with respect to the Fermi level of the metal surface.
The energy differences among these four electronic states are given by $h_0^--h_0=E_d=E_0 + \kappa_0Q_t$, $h_1-h_0=E_1 + \kappa_1Q_t$ and $h_2-h_0=E_2 + \kappa_2Q_t$, where $E_0 + \kappa_0Q_t$ is the effective LUMO energy,
$E_1$ and $E_2$ are vertical excited energies at the Franck-Condon point. The parameters $\kappa_i (i=1,2)$ represent intrastate electron-vibrational coupling constants.
In this study, we set $h_0^-=0$, assuming that the Fermi level is energetically aligned with anion state of pyrazine.
Actually, fermi level of the metal can be modulated by surface coating\cite{kim2017reduction}, doping\cite{shi2010work}, interfacial dipole engineering\cite{ma2023recent} and so on.
$\lambda$ is the $S_1-S_2$ vibronic-coupling constant.
The numerical values of the parameters for the system part Hamiltonian in Eq. \ref{Hs} are taken from Ref. \citenum{schneider1990aspects}:
$\hbar\omega_c=118meV$, $\hbar\omega_t=74meV$, $\kappa_1=-105meV$, $\kappa_2=149meV$, $\lambda=261.6meV$, $E_1=3.94eV$, $E_2=4.84eV$.

The nuclear positions and momenta are represented by $\hat{\boldsymbol{R}}$ and $\hat{\boldsymbol{P}}$, with $\alpha$ indexing the nuclear DOF. The nuclear mass is denoted by $m_{\alpha}$, and $U_0(\hat{\boldsymbol{R}})$ represents the diabatic nuclear potential corresponding to the unoccupied electronic state.
In this study, we consider three-state two-mode model of pyrazine for the system part, such that
\begin{equation}
U_0(\hat{\boldsymbol{R}}) = \sum_{\sigma}\frac{1}{2}\hbar\omega_{\sigma}Q_{\sigma}^2
\end{equation}
where $\omega_{\sigma}$ ($\sigma=t$ means tuning mode, $\sigma=c$ means coupling mode) are nuclear oscillation frequencies. 

The plasmonic nanocavity's metal states are modeled as a bath Hamiltonian $\hat{H}_B$, with $\hat{c}^+_k$ ($\hat{c}_k$) denotes the creation (annihilation) operator for the $k$-th electronic orbital of the metal bath.
The coupling between the effective molecular orbital $\hat{d}$ and the metallic orbital $\hat{c}_k$ is given by $V_{k}$. Under the wide-band approximation, the hybridization function $\Gamma(\epsilon)$, which measures the strength of the system-bath coupling, is assumed to be independent of $\epsilon$ and is expressed as:
\begin{equation}
    \Gamma(\epsilon) = 2\pi\sum_k \abs{V_{k}}^2\delta(\epsilon-\epsilon_k)=\Gamma
\end{equation}

Taking into account the light-matter interaction in a plasmonic nanocavity, the Hamiltonian becomes time-periodic. Here, we apply Floquet theory to convert the time-periodic Hamiltonian into a time-independent representation within Floquet space\cite{shirley1965solution}.
As introduced in previous works\cite{mosallanejad2023floquet}, to construct time-independent Floquet Hamiltonian $\hat{H}^F$, we define two key operators: the Fourier number operator $\hat{N}$ and the Fourier ladder operators $\hat{L}_n$. These operators exhibit the following properties:
\begin{equation}
    \hat{N}\ket{n}=n\ket{n}, \hat{L}_n\ket{m}=\ket{n+m}
\end{equation}
where $\ket{n}$ is the basis set in the Fourier space.
Then the Hamiltonian and density operator in Floquet representation would be
\begin{equation}\label{H-Floq}
    \hat{H}^F = \sum_n\hat{H}^{(n)}\hat{L}_n + \hat{N}\hbar\Omega
\end{equation}
\begin{equation}
    \hat{\rho}^F(t) = \sum_n\hat{\rho}^{(n)}(t)\hat{L}_n
\end{equation}
Here, $\hat{H}^{(n)}$ and $\hat{\rho}^{(n)}(t)$ represent the Fourier expansion coefficients in the expansions $\hat{H}(t) = \sum_n \hat{H}^{(n)} e^{in\omega t}$ and $\hat{\rho}(t) = \sum_n \hat{\rho}^{(n)}(t)e^{in\omega t}$, respectively, where $n$ is an integer ranges from $-\infty$ to $\infty$. In practical calculation, the value of $n$ must be truncated based on the ratio of the driving amplitude and the driving frequency, $\frac{A}{\hbar\Omega}$. A smaller ratio of $\frac{A}{\hbar\Omega}$ requires a smaller $n$ to achieve convergence in the calculation. 
By applying Floquet theory, the Hamiltonian is transformed into a time-independent form; however, this comes at the cost of enlarging the Hamiltonian matrix.
In the Floquet representation, considering the smallest case, the system will have three Floquet levels ($n=\pm 1, 0$), meaning the original Hamiltonian will expand to three times its original size.

\subsection{Floquet Quantum Master Equation (FQME)}
We can demonstrate that Liouville-von Neumann (LvN) equation still hold in Floquet representation as\cite{leskes2010floquet,mosallanejad2023floquet}:
\begin{equation}\label{F-EOM0}
    \frac{\partial}{\partial t}\hat{\rho}^F(t) = -\frac{i}{\hbar}[\hat{H}^F,\hat{\rho}^F(t)]
\end{equation}
Correspondingly, the Redfield equation for the molecule density operator in Floquet representation, $\hat{\rho}_S^F(t)$ reads as
\begin{equation}\label{F-EOM}
    \frac{\partial}{\partial t}\hat{\rho}_S^F(t) = -\frac{i}{\hbar}[\hat{H}_S^F,\hat{\rho}_S^F(t)] - \doublehat{\mathcal{L}}_{BS}^F\hat{\rho}_S^F(t)
\end{equation}
We refer to Eq. (\ref{F-EOM}) as the Floquet quantum master equation(FQME). Here, $\doublehat{\mathcal{L}}_{BS}^F$ is the Redfiled operator.
In diabatic representation, the Redfield operator can be expressed as:
\begin{equation}
\begin{split}
    \doublehat{\mathcal{L}}^F_{BS}\hat{\rho}_{S}^F =& 
    \frac{\Gamma}{2\hbar}\hat{d}^{F\dagger}\hat{U}\Tilde{\mathbb{D}}^F\hat{U}^{\dagger}\hat{\rho}_{S}^F(t) \\&
    + \frac{\Gamma}{2\hbar}\hat{d}^F\hat{U}\mathbb{D}^{F\dagger}\hat{U}^{\dagger}\hat{\rho}_{S}^F(t) \\&
    - \frac{\Gamma}{2\hbar}\hat{d}^{F\dagger}\hat{\rho}_{S}^F(t)\hat{U}\mathbb{D}^F\hat{U}^{\dagger} \\&
    - \frac{\Gamma}{2\hbar}\hat{d}^F\hat{\rho}_{S}^F(t)\hat{U}\Tilde{\mathbb{D}}^{F\dagger}\hat{U}^{\dagger} + \text{h.c.}
\end{split}
\end{equation}
here, $(\mathbb{D}^F)_{NM}\equiv(\hat{U}^{\dagger}\hat{d}^F\hat{U})_{NM}f(\Tilde{E}_N-\Tilde{E}_M)$, $(\mathbb{D}^{F{\dagger}})_{NM}\equiv(\hat{U}^{\dagger}\hat{d}^{F{\dagger}}\hat{U})_{NM}f(\Tilde{E}_N-\Tilde{E}_M)$, $(\Tilde{\mathbb{D}}^F)_{NM}\equiv(\hat{U}^{\dagger}\hat{d}^F\hat{U})_{NM}(1-f(\Tilde{E}_N-\Tilde{E}_M))$, $(\Tilde{\mathbb{D}}^{F{\dagger}})_{NM}\equiv(\hat{U}^{\dagger}\hat{d}^{F{\dagger}}\hat{U})_{NM}(1-f(\Tilde{E}_N-\Tilde{E}_M))$, where $\hat{U}$ and $\Tilde{E}_N$ are the eigenvectors and eigenvalues of the Floquet system Hamiltonian $\hat{H}^F_{S}$, respectively. 
$f(E)=1/(e^{E/(kT)}+1)$ is the Fermi function.
To solve the FQME, we choose 20 harmonic oscillator basis sets for each nuclear vibration mode. 

\subsection{Floquet Surface Hopping (FSH)}
Surface hopping is a hybrid method that combines classical and quantum mechanics to model molecular dynamics. In this approach, the nuclear motion is treated classically, meaning that the nuclei follow trajectories based on Newtonian mechanics, and their positions and velocities are updated at each time step. However, the electronic degrees of freedom are treated quantum mechanically, with the system being assumed to be on one of several electronic potential energy surfaces (PESs) at any given moment.
To derive surface hopping algorithm, we first perform a partial Wigner transform for Eq. \ref{F-EOM}  to approximate the nuclear motion classically. Then we arrive at a Floquet quantum-classical Liouville equation-classical master equation (FQCLE-CME) (details can be found in Ref. \citenum{wang2024nonadiabatic}):
\begin{equation}\label{FQCLE-CME}
\begin{split}
    \frac{\partial}{\partial t}\hat{\rho}_{sW}^F(\boldsymbol{R,P},t) =& \frac{1}{2}\{\hat{H}_{sW}^F(\boldsymbol{R,P}),\hat{\rho}_{sW}^F\} - \frac{1}{2}\{\hat{\rho}_{sW}^F,\hat{H}_{sW}^F(\boldsymbol{R,P})\} \\&
    - \frac{i}{\hbar}[\hat{H}_{sW}^F,\hat{\rho}_{sW}^F] - \doublehat{\mathcal{L}}^F_{bsW}(\boldsymbol{R})\hat{\rho}^F_{sW}(t)
\end{split}
\end{equation}
here, $\{\cdot,\cdot\}$ is the Poisson bracket,
\begin{equation}
    \{A,B\} = \sum_{\alpha}\left(\frac{\partial A}{\partial R_{\alpha}}\frac{\partial B}{\partial P_{\alpha}} - \frac{\partial A}{\partial P_{\alpha}}\frac{\partial B}{\partial R_{\alpha}}\right)
\end{equation}
and $\hat{H}_{sW}^F(\boldsymbol{R,P})$ indicates the partial Wigner transformation of $\hat{H}_s^F$. Here, $\doublehat{\mathcal{L}}^F_{bsW}(\boldsymbol{R})$ becomes position-dependent due to the position-dependent eigenvectors and eigenvalues of $\hat{H}_{sW}^F(\boldsymbol{R,P})$.
Typically, surface hopping is performed within the adiabatic representation. Therefore, we need FQCLE-CME in adiabatic representation:
\begin{equation}\label{FQCLE-CME-adia}
\begin{split}
    \frac{\partial}{\partial t}\rho_{NM}^{F(ad,sW)}(\boldsymbol{R,P},t) = &-\frac{i}{\hbar}(\Tilde{E}_N^{ad}(\boldsymbol{R})-\Tilde{E}_M^{ad}(\boldsymbol{R}))\rho_{NM}^{F(ad,sW)}  \\& - \sum_{\alpha K}\frac{P_{\alpha}}{m_{\alpha}}\left(d_{NK}^{\alpha}(\boldsymbol{R})\rho_{KM}^{F(ad,sW)}-\rho_{NK}^{F(ad,sW)}d_{KM}^{\alpha}(\boldsymbol{R})\right) \\& - \frac{1}{2}\sum_{\alpha K}\left(F_{NK}^{\alpha}(\boldsymbol{R})\frac{\partial\rho_{KM}^{F(ad,sW)}}{\partial P_{\alpha}} + \frac{\partial\rho_{NK}^{F(ad,sW)}}{\partial P_{\alpha}}F_{KM}^{\alpha}(\boldsymbol{R})\right) \\& - \sum_{\alpha}\frac{P_{\alpha}}{m_{\alpha}}\frac{\partial\rho_{NM}^{F(ad,sW)}}{\partial R_{\alpha}} - \sum_{KL}\mathcal{L}_{NM,KL}^{F(ad,bsW)}(\boldsymbol{R})\rho_{KL}^{F(ad,sW)}
\end{split}
\end{equation}
Here, $F^{\alpha}$ and $d^{\alpha}$ are the force and the derivative coupling.
This FQCLE-CME can be solved by Floquet surface hopping (FSH) algorithm by evolving electronic and nuclear part according to:
\begin{equation}\label{rho_dot}
\begin{split}
   \dot{\sigma}^{F(ad)}_{NM}=&
   -\frac{i}{\hbar}(\Tilde{E}_N^{ad}(\boldsymbol{R})-\Tilde{E}_M^{ad}(\boldsymbol{R}))\sigma_{NM}^{F(ad)} \\ & - \sum_{\alpha K}\frac{P_{\alpha}}{m_{\alpha}}\left(d_{NK}^{\alpha}(\boldsymbol{R})\sigma_{KM}^{F(ad)} -\sigma_{NK}^{F(ad)}d_{KM}^{\alpha}(\boldsymbol{R})\right) \\ & - \sum_{KL}\mathcal{L}_{NM,KL}^{F(ad,bsW)}(\boldsymbol{R})\sigma_{KL}^{F(ad)}
\end{split}
\end{equation}
\begin{equation}\label{Rdot}
    \dot{R}_{\alpha} = \frac{P_{\alpha}}{m_{\alpha}}
\end{equation}
\begin{equation}\label{Pdot}
    \dot{P}_{\alpha} = F_{\alpha({\lambda\lambda})}
\end{equation}
The hopping rates have two parts: one is caused by the derivative coupling ($k_{N\rightarrow M}^D$), the other one is caused by the interaction between the molecule and the metal surface ($k_{N\rightarrow M}^{\mathcal{L}}$):
\begin{align}\label{kd}
    k_{N\rightarrow M}^D = \Theta\left(-2Re\sum_{\alpha}\frac{P_{\alpha}}{m_{\alpha}}\frac{d_{MN}^{\alpha}\sigma_{NM}^{F(ad)}}{\sigma_{NN}^{F(ad)}}\right)
\end{align}
where $\Theta$ function is defined as
\begin{align}
    \Theta(x) =
    \begin{cases}
        x, (x\geq0) \\
        0, (x<0)
    \end{cases}
\end{align}
\begin{equation}\label{kL}
\begin{split}
    k_{N\rightarrow M}^{\mathcal{L}} = -\mathcal{L}_{MM,NN}^{F(ad,bsW)}
\end{split}
\end{equation}
where we employ the secular approximation to ignore the off-diagonal part of $\mathcal{L}$, 
which is accurate in long time dynamics\cite{dou2017generalized}.
The details of operating such FSH algorithm step by step are given in Ref. \citenum{wang2024nonadiabatic}.

\section{3. RESULTS AND DISCUSSION}
\begin{figure}
    \centering
    \includegraphics[width=1.0\linewidth]{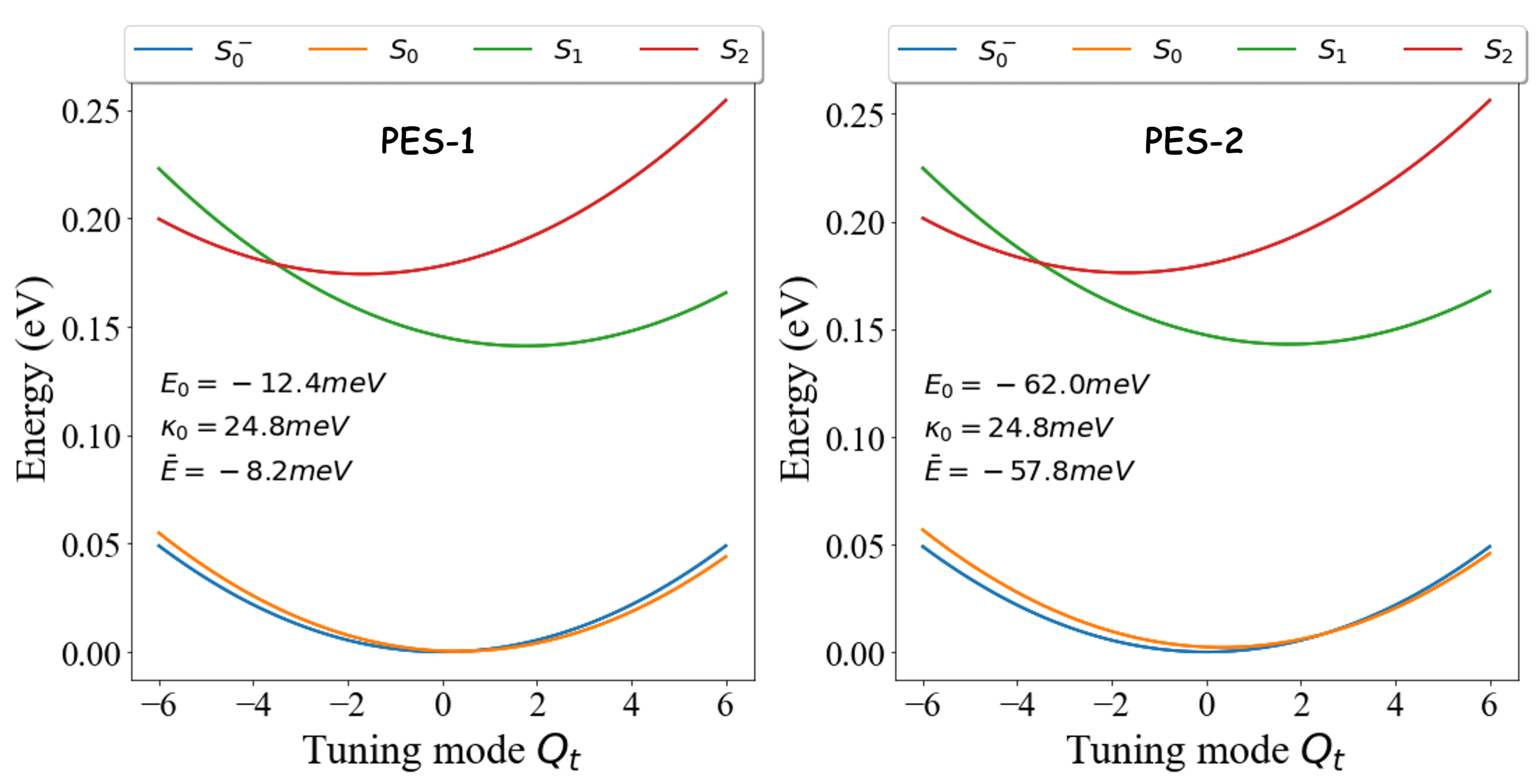}
    \caption{Diabatic potential energy surfaces (PESs) of the $S_0^-$, $S_0$, $S_1$, and $S_2$ states along the tuning mode $Q_t$. The anion of pyrazine is more stable in PES-2 of right panel. }
    \label{fig:PES-origin}
\end{figure}

In the present study, we model the potential energy surface (PES) of the pyrazine anion state $S_0^-$ based on experimental results.
Experiments indicate that the radical anion of pyrazine is generally more stable than the neutral pyrazine molecule, owing to its $\pi$-system. The delocalized $\pi$-system in pyrazine allows for the spreading of the additional electron over the ring.\cite{nenner1975temporary,song2002photoelectron,palihawadana2012electron}. Moreover, SERS data suggest that the PES of $S_0^-$ shift in the same direction as the $S_2$ state relative to the ground state along the tuning mode, but to a lesser extent\cite{arenas1996charge}.  
Accordingly, we define $\Bar{E}=\frac{\kappa_0^2}{2\hbar\omega_t}+E_0$ as the energy difference between the minimum of $S_0^-$ and $S_0$ states, and set it to be negative. And also we choose $\kappa_0$ to have the same sign as $\kappa_2$ but with a smaller absolute value than $\kappa_2$. 
Here, we define two distinct sets of PES parameters to represent two different levels of stability for the pyrazine anion. These diabatic PESs are shown in Figure \ref{fig:PES-origin}.
The pyrazine anion is more stable in PES-2 of the Figure \ref{fig:PES-origin} since the absolute value of $\Bar{E}$ is larger. In the following, we perform dynamics based on these two PESs.
For the $S_2\xrightarrow{}S_1/S_0/S_0^-$ processes, we prepare the molecule at electronic state $S_2$, assuming it is excited vertically from the minimum of the $S_0$ state. 
For the $S_0\xrightarrow{}S_0^-/S_1/S_2$ processes, we prepare the molecule at the minimum of the $S_0$ state. The molecule is then either pumped to the $S_2$ state due to the driving force, or transferred to the $S_0^-$ state due to the electron transfer from the metal surface.

\subsection{$S_2\xrightarrow{}S_1/S_0/S_0^-$ PROCESSES}
\begin{figure}
    \centering
    \includegraphics[width=1.0\linewidth]{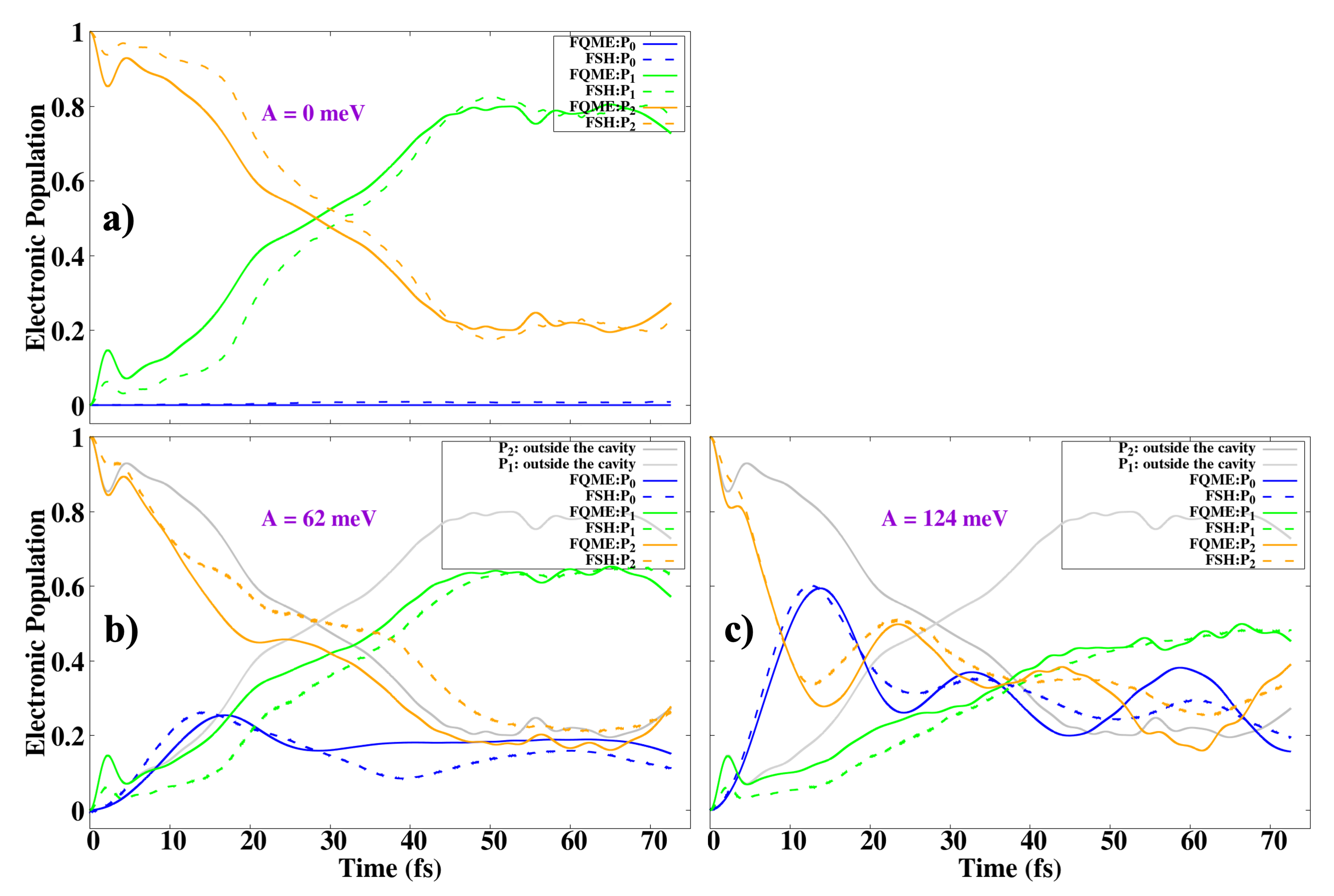}
    \caption{Population dynamics of $S_0$, $S_1$, and $S_2$ states under varying driving conditions, in the absence of electron transfer between the molecule and the metal surface ($\Gamma = 0meV$). The molecule is prepared in the $S_2$ state. The solid line represents the FQME results, while the dashed line corresponds to the FSH results. We set a resonant driving frequency of $\hbar\Omega=\Delta E_{02}$. a) The driving amplitude $A=0 meV$, b) the driving amplitude $A=62 meV$, and c) the driving amplitude $A=124 meV$. We observe good agreement between FQME and FSH, and the electronic relaxation on $S_2$ state speeds up as the driving amplitude increases. We observe good agreement between the FQME and FSH results.}
    \label{fig:init_s2_no_Gamma}
\end{figure}

First, we consider the scenario where no electron transfer occurs between the molecule and the metal surface, with the molecule initialized on the $S_2$ state.
In this case, the $S_0^-$ state will not be occupied.
When the molecule is outside the plasmonic cavity, where no external driving force acts on the molecule, the electronic state population dynamics are depicted in Figure \ref{fig:init_s2_no_Gamma}a.
The electronic excitation relaxes to the $S_1$ state within 30 $fs$ due to the conical intersection (CI) and the strong electron-nuclear coupling\cite{gu2020manipulating}.
When the molecule is inside the plasmonic nanocavity, the electromagnetic field couples with the transition dipole moment between $S_0$ and $S_2$ states, facilitating electronic state population transfer between these two states. 
Here, we apply a resonant driving force, with the driving frequency matching the energy difference between the $S_2$ and $S_0$ states ($\hbar\Omega=\Delta E_{02}$).
Under these conditions, there are two possible channels for the relaxation of electronic excitation from the $S_2$ state ($S_2\xrightarrow{}S_1$ and $S_2\xrightarrow{}S_0$), leading to a faster relaxation dynamics in the $S_2$ state (within 25 $fs$), as illustrated in Figure \ref{fig:init_s2_no_Gamma}b.
As the driving strength $A$ increases, 
populations in the $S_2$ state relax to both $S_0$ and $S_1$ states within 15 $fs$, as shown in Figure \ref{fig:init_s2_no_Gamma}c.
At early times, the population in the $S_0$ state is significantly larger than that in the $S_1$ state due to the strong driving force. Moreover, 
the oscillation of populations between $S_0$ and $S_2$ states are observed, driven by the intense external field.
It is noteworthy that the FQME and FSH methods agree closely with each other.

The rapid relaxation caused by the driving forces can also be understood in terms of nuclear dynamics. Figure S2 illustrates the nuclear dynamics along the tuning mode under different driving conditions, assuming no electron transfer between the molecule and the metal surface. Outside the cavity, the nuclei pass through the CI. However, inside the cavity, the electron can relax via the avoided crossing point induced by the driving, as shown in the Floquet PES in Figure S3. In the Floquet PES, the original PES for each state is expanded into multiple Floquet replicas according to Eq. \ref{H-Floq}, with the energy difference between each adjacent Floquet replicas corresponding to the driving frequency. Avoided crossings occur between Floquet replicas of $S_0$ and $S_2$ states in the adiabatic representation. As a result, during the relaxation processes, the nuclear motion along the tuning mode can also pass through avoided crossing points, and does not necessarily need to reach the CI.

\begin{figure}
    \centering
    \includegraphics[width=1.0\linewidth]{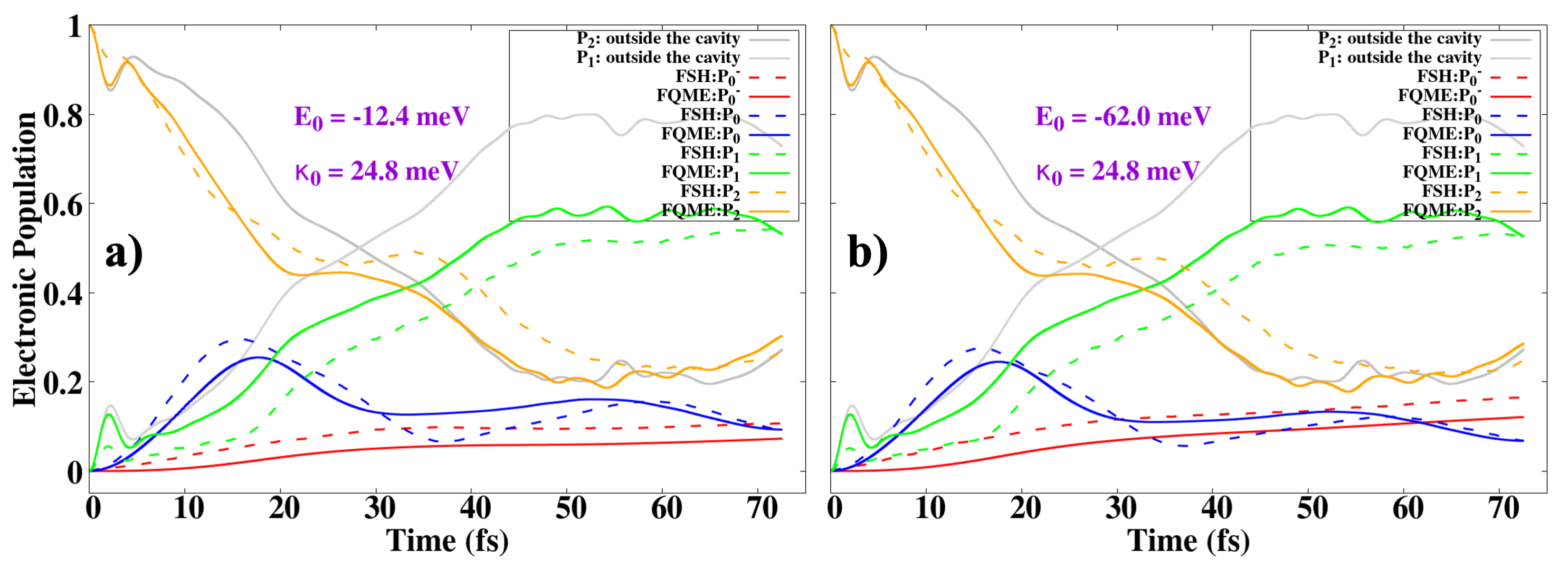}
    \caption{Population dynamics of $S_0^-$, $S_0$, $S_1$, and $S_2$ states under a weak driving force of $A=62meV$. The molecule is prepared in the $S_2$ state. We consider two different PESs of $S_0^-$, each characterized by different values of $E_0$ and $\kappa_0$ (see Figure \ref{fig:PES-origin}). We set a weak molecule-metal coupling of $\Gamma=12.4meV$. The solid line represents the FQME results, while the dashed line corresponds to the FSH results. We set a resonant driving frequency of $\hbar\Omega=\Delta E_{02}$. Electronic dynamics a) on PES-1, and b) on PES-2. We observe good agreement between the FQME and FSH results.}
    \label{fig:init_s2_Gamma_100_A500}
\end{figure}

\begin{figure}
    \centering
    \includegraphics[width=1.0\linewidth]{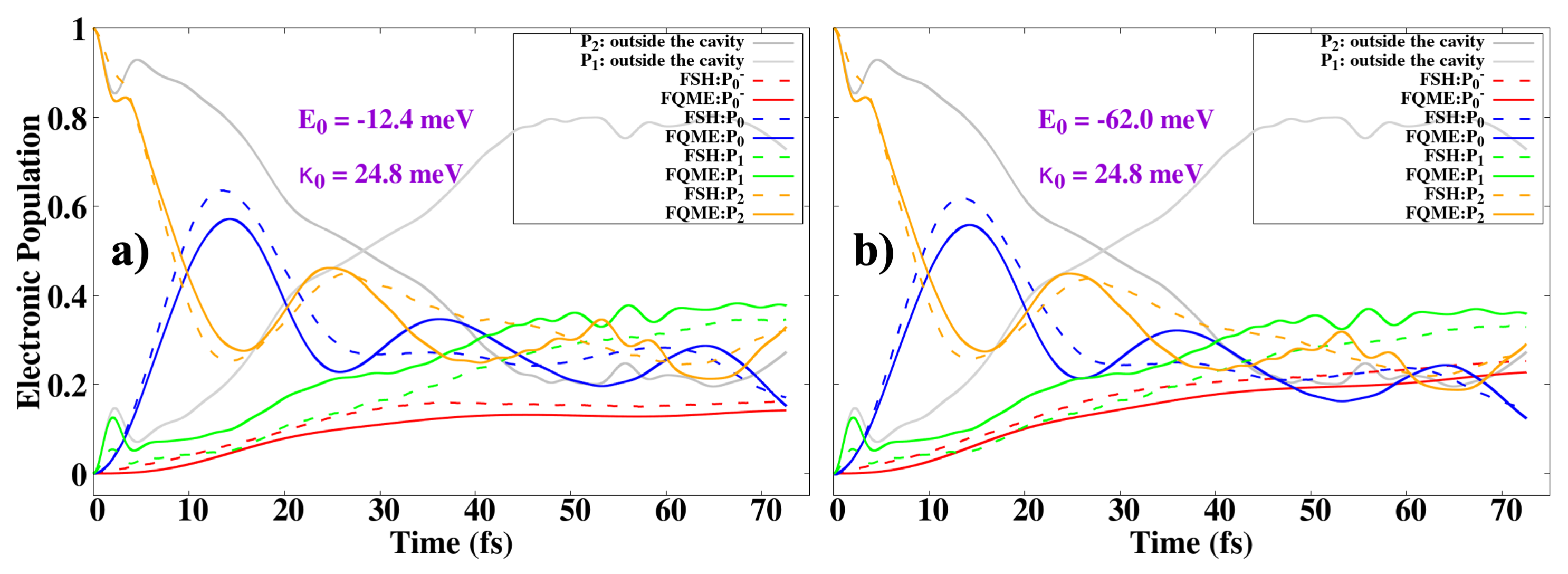}
    \caption{Population dynamics of $S_0^-$, $S_0$, $S_1$, and $S_2$ states under a strong driving force of $A=124meV$. The molecule is prepared in the $S_2$ state. We consider two different PESs of $S_0^-$, each characterized by different values of $E_0$ and $\kappa_0$ (see Figure \ref{fig:PES-origin}). We set a weak molecule-metal coupling of $\Gamma=12.4meV$. The solid line represents the FQME results, while the dashed line corresponds to the FSH results. We set a resonant driving frequency of $\hbar\Omega=\Delta E_{02}$. Electronic dynamics a) on PES-1, and b) on PES-2. We observe good agreement between the FQME and FSH results.}
    \label{fig:init_s2_Gamma_100_A1000}
\end{figure}

Next, we consider the scenario where electron transfer occurs between the molecule and the metal surface as shown in Figure \ref{fig:init_s2_Gamma_100_A500}.
Here, we set a weak coupling between the molecule and the metal surface as $\Gamma=12.4meV$.
In Figure \ref{fig:init_s2_Gamma_100_A500}, we show the electronic states population dynamics when adding a weak driving of $A=62meV$. In this case, $S_0^-$ state begins to populate as a function of time. Comparing with Figure \ref{fig:init_s2_no_Gamma}b, we find that 1) the population of the $S_0$ decreases, because the system gains electrons from metal surface when it is in the $S_0$ state, leading to a transition to the $S_0^-$ state; 2) The population of the $S_1$ decreases because the molecule tends to decay nonadiabtically through avoided crossing points to the $S_0$ state; 3) The population of the $S_0^-$ state increases more rapidly with time in Figure \ref{fig:init_s2_Gamma_100_A500}b than in Figure \ref{fig:init_s2_Gamma_100_A500}a, indicating that the electron transfer rate between the molecule and the metal surface is higher on PES-2, where the $S_0^-$ state is more stable; 4) The electron transfer between the molecule and the metal surface does not affect the relaxation rate.
Figure \ref{fig:init_s2_Gamma_100_A1000}
shows the  population dynamics when a stronger driving of $A=124meV$ is applied.
Comparing this with Figure \ref{fig:init_s2_no_Gamma}c, we observe similar changes in the dynamics as seen with the weaker driving field. Additionally, under the influence of a stronger external field, the population of the $S_0^-$ state increases more rapidly over time compared to Figure \ref{fig:init_s2_Gamma_100_A500}.
This is because the population of the $S_0^-$ state is influenced not only by the coupling strength between the molecule and the metal surface, but also by the population of the $S_0$ state. Stronger fields accelerate the population of the $S_0$ state, leading to a larger population. This, in turn, enhances electron transfer processes between the molecule and the metal surface under these conditions.
We observe good agreement between the FQME and FSH results in 
Figures \ref{fig:init_s2_Gamma_100_A500} and \ref{fig:init_s2_Gamma_100_A1000}.

In this section, we investigate the population dynamics of a molecule initialized in the $S_2$ state under various driving conditions, considering both cases of electron transfer and no electron transfer between the molecule and the metal surface.
We observe that 1) the plasmonic cavity effectively accelerates relaxation from the $S_2$ state and enhances electron transfer between the molecule and the metal surface; 2) the electron transfer processes affect the state populations but do not further impact the overall relaxation dynamics rate beyond the effect of the driving.

\subsection{$S_0\xrightarrow{}S_0^-/S_1/S_2$ PROCESSES}
\begin{figure}
    \centering
    \includegraphics[width=1.0\linewidth]{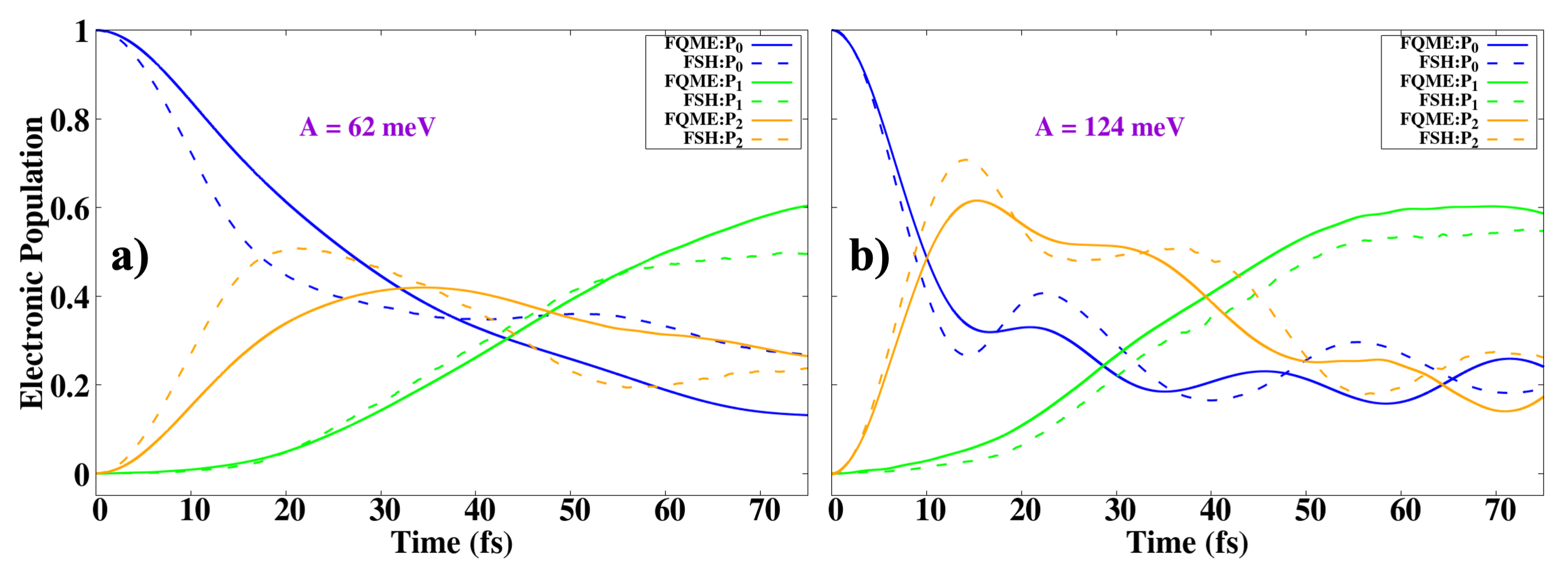}
    \caption{Population dynamics of $S_0$, $S_1$, and $S_2$ states under varying driving conditions, in the absence of electron transfer between the molecule and the metal surface ($\Gamma = 0meV$). The molecule is prepared in the $S_0$ state. The solid line represents the FQME results, while the dashed line corresponds to the FSH results. We set a resonant driving frequency of $\hbar\Omega=\Delta E_{02}$. a) the driving amplitude $A=62 meV$, and b) the driving amplitude $A=124 meV$. We observe good agreement between the FQME and FSH results.}
    \label{fig:init_s0_no_Gamma}
\end{figure}

For these processes, we firstly consider the scenario where no electron transfer occurs between the molecule and the metal surface, with the molecule prepared in the $S_0$ state. In this case, a driving force must be applied to the system, otherwise, the molecule will remain in the $S_0$ state. When a weaker driving force is applied, the molecule is first pumped to the $S_2$ state, and then relaxes to the $S_1$ state from the $S_2$, as shown in Figure \ref{fig:init_s0_no_Gamma}a.
When a stronger driving force is applied, the molecule populates the $S_2$ state more rapidly and to a greater extent, as shown in Figure \ref{fig:init_s0_no_Gamma}b.
Moreover, the relaxation from the $S_2$ state is faster under the stronger driving force.

\begin{figure}
    \centering
    \includegraphics[width=1.0\linewidth]{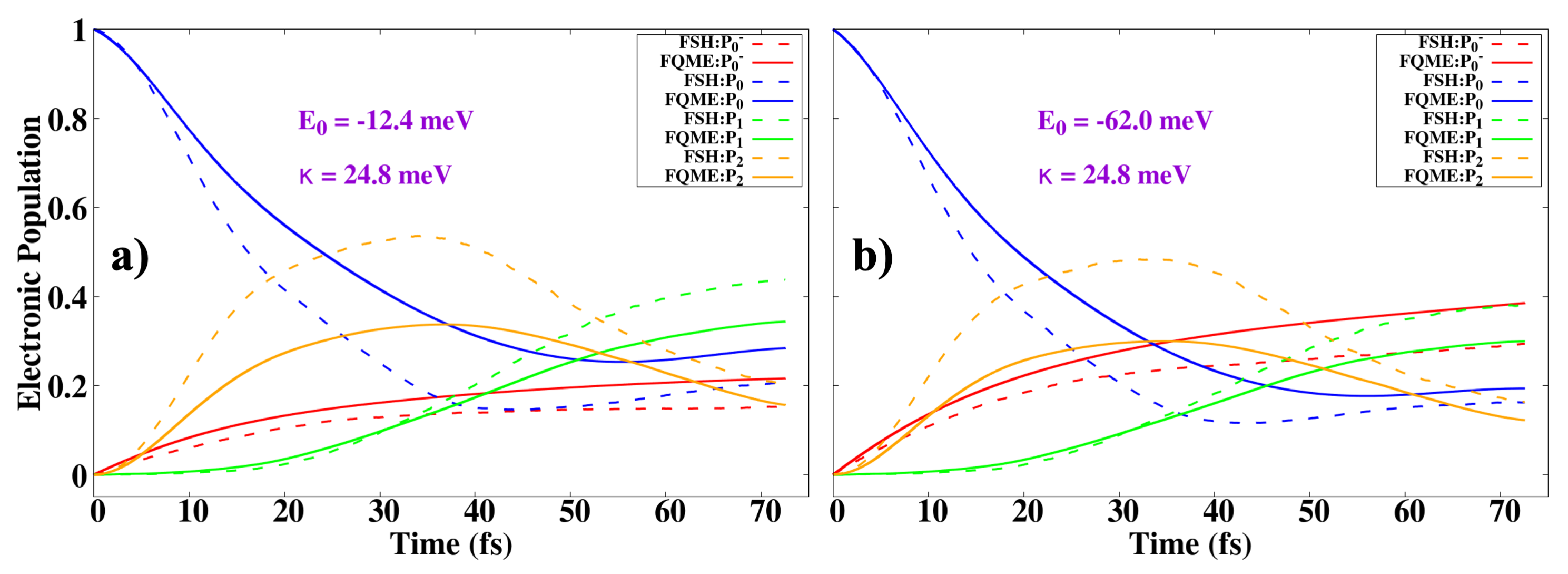}
    \caption{Population dynamics of $S_0^-$, $S_0$, $S_1$, and $S_2$ states under a weak driving force of $A=62meV$. The molecule is prepared in the $S_0$ state. We consider two different PESs of $S_0^-$, each characterized by different values of $E_0$ and $\kappa_0$ (see Figure \ref{fig:PES-origin}). We set a weak molecule-metal coupling of $\Gamma=12.4meV$. The solid line represents the FQME results, while the dashed line corresponds to the FSH results. We set a resonant driving frequency of $\hbar\Omega=\Delta E_{02}$. Electronic dynamics a) on PES-1, and b) on PES-2. We observe good agreement between the FQME and FSH results.}
    \label{fig:init_s0_Gamma_100_A500}
\end{figure}

\begin{figure}
    \centering
    \includegraphics[width=1.0\linewidth]{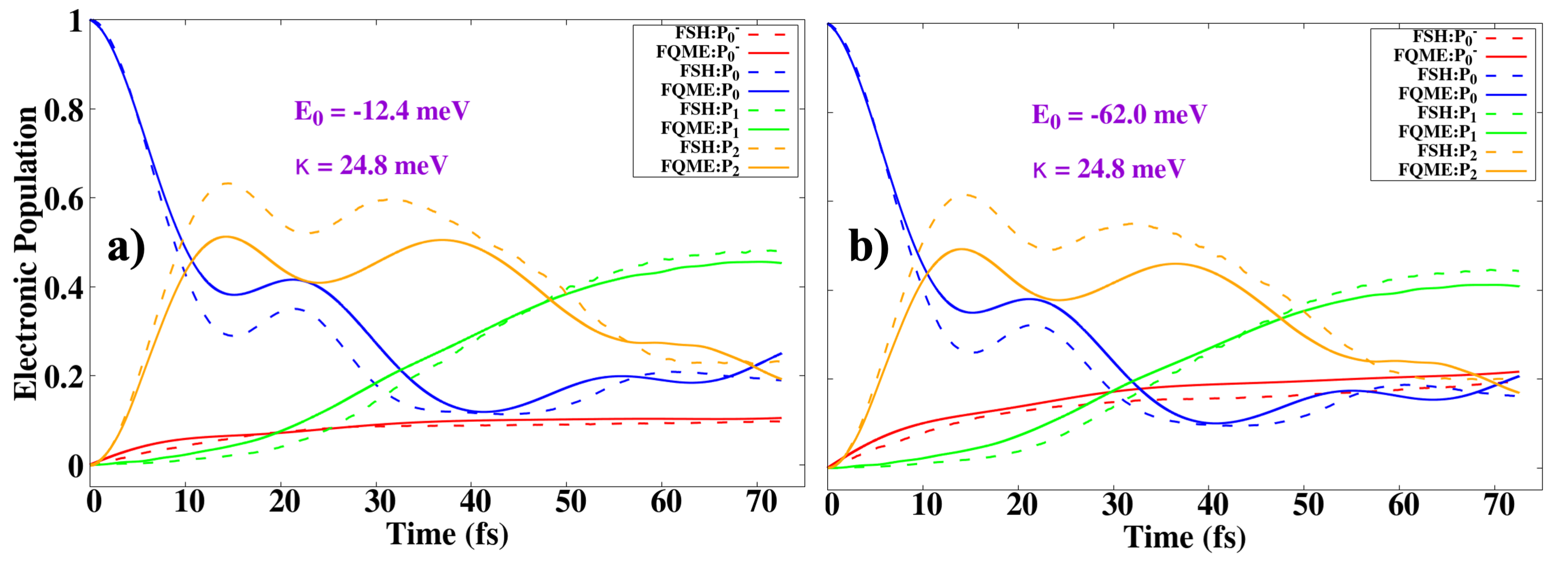}
    \caption{Population dynamics of $S_0^-$, $S_0$, $S_1$, and $S_2$ states under a weak driving force of $A=124meV$. The molecule is prepared in the $S_0$ state. We consider two different PESs of $S_0^-$, each characterized by different values of $E_0$ and $\kappa_0$ (see Figure \ref{fig:PES-origin}). We set a weak molecule-metal coupling of $\Gamma=12.4meV$. The solid line represents the FQME results, while the dashed line corresponds to the FSH results. We set a resonant driving frequency of $\hbar\Omega=\Delta E_{02}$. Electronic dynamics a) on PES-1, and b) on PES-2. We observe good agreement between the FQME and FSH results.}
    \label{fig:init_s0_Gamma_100_A1000}
\end{figure}

Next, we consider the scenario in which electron transfer takes place between the molecule and the metal surface, and we consider a weak driving force, as depicted in Figure \ref{fig:init_s0_Gamma_100_A500}.
In this case, molecules in the $S_0$ state can be either transferred to the $S_2$ state, or to the $S_0^-$ state at early times.
As a result, compared to Figure \ref{fig:init_s0_no_Gamma}a, the population in the $S_2$ state decreases, while the population in the $S_0^-$ increases. 
In Figure \ref{fig:init_s0_Gamma_100_A500}b, the population of the $S_0^-$ state increases more rapidly than in Figure \ref{fig:init_s0_Gamma_100_A500}a, indicating a faster electron transfer between the molecule and the metal surface. This is because the $S_0^-$ state is more stable under PES-2 condition.
When a strong driving force is applied on the system, the population of the $S_0^-$ state also increases, as shown in Figure \ref{fig:init_s0_Gamma_100_A1000}. However, the increase is less pronounced than with a weak driving force.
This is because the strong driving force facilitates the transition from the $S_0$ state to the $S_2$ state. Since the coupling between the molecule and the metal surface is weak, the electron transfer processes occur much more slowly than the pumping process.
Note that when the molecules are initially prepared in the $S_0$ state, the driving effect on electron transfer processes differs from that observed when they are prepared in the $S_2$ state. This is because electron transfer processes are influenced by the population of the $S_0$ state.

In this section, we investigate the population dynamics of a molecule initialized in the $S_0$ state, examining both scenarios with and without electron transfer between the molecule and the metal surface under different driving conditions.
We observe that 1) the plasmonic cavity effectively facilitates both pumping and relaxation processes; 2) while electron transfer processes influence state populations, they do not further impact the overall rates of pumping and relaxation dynamics beyond the effect of the driving.

\section{4. CONCLUSIONS}
In this study, we used pyrazine as a model molecule to investigate the influence of plasmonic nanocavities on nonadiabatic dynamics. Our findings reveal that when the molecule is initialized in the $S_2$ state, the plasmonic cavity effectively accelerates relaxation from the $S_2$ state and enhances electron transfer between the molecule and the metal surface. Although these electron transfer processes affect the state populations, they do not significantly alter the overall relaxation rate beyond the effect of the external driving field. In contrast, when the molecule is initialized in the $S_0$ state, the plasmonic cavity facilitates both pumping and relaxation processes. Here, electron transfer processes influence state populations but do not further impact the overall rates of pumping and relaxation beyond the driving effect. These results highlight the role of plasmonic cavities in modulating molecular dynamics, offering insights into how plasmonic interactions can be tailored to control electron and energy transfer in ultrafast processes. 
FSH method, which is computationally less demanding, agrees well with FQME under all the conditions we considered.
Therefore, we can rely on the FSH method for simulating larger systems in the future.
Our work opens up new possibilities for designing plasmonic systems that can manipulate both electronic transitions and relaxation dynamics in a variety of applications, from chemical reactions to quantum information processing.

\begin{acknowledgement}

D.W. acknowledges supports from the National Natural
Science Foundation of China No. 22273075. Y.W. acknowledges supports from the National Natural Science Foundation of China No. 22403077.

\end{acknowledgement}

\bibliography{acs}

\providecommand{\latin}[1]{#1}
\makeatletter
\providecommand{\doi}
  {\begingroup\let\do\@makeother\dospecials
  \catcode`\{=1 \catcode`\}=2 \doi@aux}
\providecommand{\doi@aux}[1]{\endgroup\texttt{#1}}
\makeatother
\providecommand*\mcitethebibliography{\thebibliography}
\csname @ifundefined\endcsname{endmcitethebibliography}
  {\let\endmcitethebibliography\endthebibliography}{}
\begin{mcitethebibliography}{47}
\providecommand*\natexlab[1]{#1}
\providecommand*\mciteSetBstSublistMode[1]{}
\providecommand*\mciteSetBstMaxWidthForm[2]{}
\providecommand*\mciteBstWouldAddEndPuncttrue
  {\def\EndOfBibitem{\unskip.}}
\providecommand*\mciteBstWouldAddEndPunctfalse
  {\let\EndOfBibitem\relax}
\providecommand*\mciteSetBstMidEndSepPunct[3]{}
\providecommand*\mciteSetBstSublistLabelBeginEnd[3]{}
\providecommand*\EndOfBibitem{}
\mciteSetBstSublistMode{f}
\mciteSetBstMaxWidthForm{subitem}{(\alph{mcitesubitemcount})}
\mciteSetBstSublistLabelBeginEnd
  {\mcitemaxwidthsubitemform\space}
  {\relax}
  {\relax}

\bibitem[Kwon \latin{et~al.}(2014)Kwon, No, and Park]{kwon2014design}
Kwon,~S.-H.; No,~Y.-S.; Park,~H.-G. Design of plasmonic cavities. \emph{Nano
  Convergence} \textbf{2014}, \emph{1}, 1--9\relax
\mciteBstWouldAddEndPuncttrue
\mciteSetBstMidEndSepPunct{\mcitedefaultmidpunct}
{\mcitedefaultendpunct}{\mcitedefaultseppunct}\relax
\EndOfBibitem
\bibitem[Hugall \latin{et~al.}(2018)Hugall, Singh, and van
  Hulst]{hugall2018plasmonic}
Hugall,~J.~T.; Singh,~A.; van Hulst,~N.~F. Plasmonic cavity coupling. \emph{Acs
  Photonics} \textbf{2018}, \emph{5}, 43--53\relax
\mciteBstWouldAddEndPuncttrue
\mciteSetBstMidEndSepPunct{\mcitedefaultmidpunct}
{\mcitedefaultendpunct}{\mcitedefaultseppunct}\relax
\EndOfBibitem
\bibitem[Bitton and Haran(2022)Bitton, and Haran]{bitton2022plasmonic}
Bitton,~O.; Haran,~G. Plasmonic cavities and individual quantum emitters in the
  strong coupling limit. \emph{Accounts of chemical research} \textbf{2022},
  \emph{55}, 1659--1668\relax
\mciteBstWouldAddEndPuncttrue
\mciteSetBstMidEndSepPunct{\mcitedefaultmidpunct}
{\mcitedefaultendpunct}{\mcitedefaultseppunct}\relax
\EndOfBibitem
\bibitem[Mandal \latin{et~al.}(2023)Mandal, Taylor, Weight, Koessler, Li, and
  Huo]{mandal2023theoretical}
Mandal,~A.; Taylor,~M.~A.; Weight,~B.~M.; Koessler,~E.~R.; Li,~X.; Huo,~P.
  Theoretical advances in polariton chemistry and molecular cavity quantum
  electrodynamics. \emph{Chemical Reviews} \textbf{2023}, \emph{123},
  9786--9879\relax
\mciteBstWouldAddEndPuncttrue
\mciteSetBstMidEndSepPunct{\mcitedefaultmidpunct}
{\mcitedefaultendpunct}{\mcitedefaultseppunct}\relax
\EndOfBibitem
\bibitem[Ditlbacher \latin{et~al.}(2005)Ditlbacher, Hohenau, Wagner, Kreibig,
  Rogers, Hofer, Aussenegg, and Krenn]{ditlbacher2005silver}
Ditlbacher,~H.; Hohenau,~A.; Wagner,~D.; Kreibig,~U.; Rogers,~M.; Hofer,~F.;
  Aussenegg,~F.~R.; Krenn,~J.~R. Silver nanowires as surface plasmon
  resonators. \emph{Physical review letters} \textbf{2005}, \emph{95},
  257403\relax
\mciteBstWouldAddEndPuncttrue
\mciteSetBstMidEndSepPunct{\mcitedefaultmidpunct}
{\mcitedefaultendpunct}{\mcitedefaultseppunct}\relax
\EndOfBibitem
\bibitem[Kress \latin{et~al.}(2015)Kress, Antolinez, Richner, Jayanti, Kim,
  Prins, Riedinger, Fischer, Meyer, McPeak, \latin{et~al.}
  others]{kress2015wedge}
Kress,~S.~J.; Antolinez,~F.~V.; Richner,~P.; Jayanti,~S.~V.; Kim,~D.~K.;
  Prins,~F.; Riedinger,~A.; Fischer,~M.~P.; Meyer,~S.; McPeak,~K.~M.,
  \latin{et~al.}  Wedge waveguides and resonators for quantum plasmonics.
  \emph{Nano letters} \textbf{2015}, \emph{15}, 6267--6275\relax
\mciteBstWouldAddEndPuncttrue
\mciteSetBstMidEndSepPunct{\mcitedefaultmidpunct}
{\mcitedefaultendpunct}{\mcitedefaultseppunct}\relax
\EndOfBibitem
\bibitem[Perney \latin{et~al.}(2007)Perney, Garc{\'\i}a~de Abajo, Baumberg,
  Tang, Netti, Charlton, and Zoorob]{perney2007tuning}
Perney,~N.; Garc{\'\i}a~de Abajo,~F.~J.; Baumberg,~J.; Tang,~A.; Netti,~M.;
  Charlton,~M.; Zoorob,~M. Tuning localized plasmon cavities for optimized
  surface-enhanced Raman scattering. \emph{Physical Review B—Condensed Matter
  and Materials Physics} \textbf{2007}, \emph{76}, 035426\relax
\mciteBstWouldAddEndPuncttrue
\mciteSetBstMidEndSepPunct{\mcitedefaultmidpunct}
{\mcitedefaultendpunct}{\mcitedefaultseppunct}\relax
\EndOfBibitem
\bibitem[Novotny and Van~Hulst(2011)Novotny, and
  Van~Hulst]{novotny2011antennas}
Novotny,~L.; Van~Hulst,~N. Antennas for light. \emph{Nature photonics}
  \textbf{2011}, \emph{5}, 83--90\relax
\mciteBstWouldAddEndPuncttrue
\mciteSetBstMidEndSepPunct{\mcitedefaultmidpunct}
{\mcitedefaultendpunct}{\mcitedefaultseppunct}\relax
\EndOfBibitem
\bibitem[Maccaferri \latin{et~al.}(2021)Maccaferri, Barbillon, Koya, Lu, Acuna,
  and Garoli]{maccaferri2021recent}
Maccaferri,~N.; Barbillon,~G.; Koya,~A.~N.; Lu,~G.; Acuna,~G.~P.; Garoli,~D.
  Recent advances in plasmonic nanocavities for single-molecule spectroscopy.
  \emph{Nanoscale Advances} \textbf{2021}, \emph{3}, 633--642\relax
\mciteBstWouldAddEndPuncttrue
\mciteSetBstMidEndSepPunct{\mcitedefaultmidpunct}
{\mcitedefaultendpunct}{\mcitedefaultseppunct}\relax
\EndOfBibitem
\bibitem[Lyu \latin{et~al.}(2023)Lyu, Liu, Yin, Wu, Sun, Chen, Xu, and
  Kang]{lyu2023periodic}
Lyu,~P.-T.; Liu,~X.-R.; Yin,~L.-X.; Wu,~P.; Sun,~C.; Chen,~H.-Y.; Xu,~J.-J.;
  Kang,~B. Periodic distributions and ultrafast dynamics of hot electrons in
  plasmonic resonators. \emph{Nano Letters} \textbf{2023}, \emph{23},
  2269--2276\relax
\mciteBstWouldAddEndPuncttrue
\mciteSetBstMidEndSepPunct{\mcitedefaultmidpunct}
{\mcitedefaultendpunct}{\mcitedefaultseppunct}\relax
\EndOfBibitem
\bibitem[Lyu \latin{et~al.}(2023)Lyu, Yin, Shen, Gao, Chen, Xu, and
  Kang]{lyu2023plasmonic}
Lyu,~P.-T.; Yin,~L.-X.; Shen,~Y.-T.; Gao,~Z.; Chen,~H.-Y.; Xu,~J.-J.; Kang,~B.
  Plasmonic cavity-catalysis by standing hot carrier waves. \emph{Journal of
  the American Chemical Society} \textbf{2023}, \emph{145}, 18912--18919\relax
\mciteBstWouldAddEndPuncttrue
\mciteSetBstMidEndSepPunct{\mcitedefaultmidpunct}
{\mcitedefaultendpunct}{\mcitedefaultseppunct}\relax
\EndOfBibitem
\bibitem[Im \latin{et~al.}(2013)Im, Bantz, Lee, Johnson, Haynes, and
  Oh]{im2013self}
Im,~H.; Bantz,~K.~C.; Lee,~S.~H.; Johnson,~T.~W.; Haynes,~C.~L.; Oh,~S.-H.
  Self-assembled plasmonic nanoring cavity arrays for SERS and LSPR biosensing.
  \emph{Advanced Materials} \textbf{2013}, \emph{25}, 2678--2685\relax
\mciteBstWouldAddEndPuncttrue
\mciteSetBstMidEndSepPunct{\mcitedefaultmidpunct}
{\mcitedefaultendpunct}{\mcitedefaultseppunct}\relax
\EndOfBibitem
\bibitem[Maier(2006)]{maier2006plasmonic}
Maier,~S.~A. Plasmonic field enhancement and SERS in the effective mode volume
  picture. \emph{Optics Express} \textbf{2006}, \emph{14}, 1957--1964\relax
\mciteBstWouldAddEndPuncttrue
\mciteSetBstMidEndSepPunct{\mcitedefaultmidpunct}
{\mcitedefaultendpunct}{\mcitedefaultseppunct}\relax
\EndOfBibitem
\bibitem[Li \latin{et~al.}(2017)Li, Feng, Teng, and Lu]{li2017fabrication}
Li,~N.; Feng,~L.; Teng,~F.; Lu,~N. Fabrication of plasmonic cavity arrays for
  SERS analysis. \emph{Nanotechnology} \textbf{2017}, \emph{28}, 185301\relax
\mciteBstWouldAddEndPuncttrue
\mciteSetBstMidEndSepPunct{\mcitedefaultmidpunct}
{\mcitedefaultendpunct}{\mcitedefaultseppunct}\relax
\EndOfBibitem
\bibitem[Sharma and Dhawan(2014)Sharma, and Dhawan]{sharma2014hybrid}
Sharma,~Y.; Dhawan,~A. Hybrid nanoparticle--nanoline plasmonic cavities as SERS
  substrates with gap-controlled enhancements and resonances.
  \emph{Nanotechnology} \textbf{2014}, \emph{25}, 085202\relax
\mciteBstWouldAddEndPuncttrue
\mciteSetBstMidEndSepPunct{\mcitedefaultmidpunct}
{\mcitedefaultendpunct}{\mcitedefaultseppunct}\relax
\EndOfBibitem
\bibitem[Ameling \latin{et~al.}(2010)Ameling, Langguth, Hentschel, Mesch,
  Braun, and Giessen]{ameling2010cavity}
Ameling,~R.; Langguth,~L.; Hentschel,~M.; Mesch,~M.; Braun,~P.~V.; Giessen,~H.
  Cavity-enhanced localized plasmon resonance sensing. \emph{Applied Physics
  Letters} \textbf{2010}, \emph{97}\relax
\mciteBstWouldAddEndPuncttrue
\mciteSetBstMidEndSepPunct{\mcitedefaultmidpunct}
{\mcitedefaultendpunct}{\mcitedefaultseppunct}\relax
\EndOfBibitem
\bibitem[Li \latin{et~al.}(2020)Li, Duan, Yi, Wang, Radjenovic, Tian, and
  Li]{li2020real}
Li,~C.-Y.; Duan,~S.; Yi,~J.; Wang,~C.; Radjenovic,~P.~M.; Tian,~Z.-Q.;
  Li,~J.-F. Real-time detection of single-molecule reaction by plasmon-enhanced
  spectroscopy. \emph{Science Advances} \textbf{2020}, \emph{6}, eaba6012\relax
\mciteBstWouldAddEndPuncttrue
\mciteSetBstMidEndSepPunct{\mcitedefaultmidpunct}
{\mcitedefaultendpunct}{\mcitedefaultseppunct}\relax
\EndOfBibitem
\bibitem[Kwon \latin{et~al.}(2018)Kwon, Jin, Shin, Kim, Kim, Kang, and
  Choi]{kwon2018tunable}
Kwon,~J.~A.; Jin,~C.~M.; Shin,~Y.; Kim,~H.~Y.; Kim,~Y.; Kang,~T.; Choi,~I.
  Tunable plasmonic cavity for label-free detection of small molecules.
  \emph{ACS applied materials \& interfaces} \textbf{2018}, \emph{10},
  13226--13235\relax
\mciteBstWouldAddEndPuncttrue
\mciteSetBstMidEndSepPunct{\mcitedefaultmidpunct}
{\mcitedefaultendpunct}{\mcitedefaultseppunct}\relax
\EndOfBibitem
\bibitem[Fojt \latin{et~al.}(2024)Fojt, Erhart, and
  Sch{"a}fer]{fojt2024controlling}
Fojt,~J.; Erhart,~P.; Sch{"a}fer,~C. Controlling Plasmonic Catalysis via Strong
  Coupling with Electromagnetic Resonators. \emph{Nano Letters} \textbf{2024},
  \emph{24}, 11913--11920\relax
\mciteBstWouldAddEndPuncttrue
\mciteSetBstMidEndSepPunct{\mcitedefaultmidpunct}
{\mcitedefaultendpunct}{\mcitedefaultseppunct}\relax
\EndOfBibitem
\bibitem[Marquier \latin{et~al.}(2017)Marquier, Sauvan, and
  Greffet]{marquier2017revisiting}
Marquier,~F.; Sauvan,~C.; Greffet,~J.-J. Revisiting quantum optics with surface
  plasmons and plasmonic resonators. \emph{ACS photonics} \textbf{2017},
  \emph{4}, 2091--2101\relax
\mciteBstWouldAddEndPuncttrue
\mciteSetBstMidEndSepPunct{\mcitedefaultmidpunct}
{\mcitedefaultendpunct}{\mcitedefaultseppunct}\relax
\EndOfBibitem
\bibitem[Fryett \latin{et~al.}(2017)Fryett, Zhan, and
  Majumdar]{fryett2017cavity}
Fryett,~T.; Zhan,~A.; Majumdar,~A. Cavity nonlinear optics with layered
  materials. \emph{Nanophotonics} \textbf{2017}, \emph{7}, 355--370\relax
\mciteBstWouldAddEndPuncttrue
\mciteSetBstMidEndSepPunct{\mcitedefaultmidpunct}
{\mcitedefaultendpunct}{\mcitedefaultseppunct}\relax
\EndOfBibitem
\bibitem[Panoiu \latin{et~al.}(2018)Panoiu, Sha, Lei, and
  Li]{panoiu2018nonlinear}
Panoiu,~N.~C.; Sha,~W.~E.; Lei,~D.; Li,~G. Nonlinear optics in plasmonic
  nanostructures. \emph{Journal of Optics} \textbf{2018}, \emph{20},
  083001\relax
\mciteBstWouldAddEndPuncttrue
\mciteSetBstMidEndSepPunct{\mcitedefaultmidpunct}
{\mcitedefaultendpunct}{\mcitedefaultseppunct}\relax
\EndOfBibitem
\bibitem[Russell \latin{et~al.}(2012)Russell, Liu, Cui, and
  Hu]{russell2012large}
Russell,~K.~J.; Liu,~T.-L.; Cui,~S.; Hu,~E.~L. Large spontaneous emission
  enhancement in plasmonic nanocavities. \emph{Nature Photonics} \textbf{2012},
  \emph{6}, 459--462\relax
\mciteBstWouldAddEndPuncttrue
\mciteSetBstMidEndSepPunct{\mcitedefaultmidpunct}
{\mcitedefaultendpunct}{\mcitedefaultseppunct}\relax
\EndOfBibitem
\bibitem[Butt(2024)]{butt2024review}
Butt,~M.~A. Review of Innovative Cavity Designs in Metal--Insulator-Metal
  Waveguide-Based Plasmonic Sensors. \emph{Plasmonics} \textbf{2024},
  1--20\relax
\mciteBstWouldAddEndPuncttrue
\mciteSetBstMidEndSepPunct{\mcitedefaultmidpunct}
{\mcitedefaultendpunct}{\mcitedefaultseppunct}\relax
\EndOfBibitem
\bibitem[Climent \latin{et~al.}(2019)Climent, Galego, Garcia-Vidal, and
  Feist]{climent2019plasmonic}
Climent,~C.; Galego,~J.; Garcia-Vidal,~F.~J.; Feist,~J. Plasmonic Nanocavities
  Enable Self-Induced Electrostatic Catalysis. \emph{Angewandte Chemie
  International Edition} \textbf{2019}, \emph{58}, 8698--8702\relax
\mciteBstWouldAddEndPuncttrue
\mciteSetBstMidEndSepPunct{\mcitedefaultmidpunct}
{\mcitedefaultendpunct}{\mcitedefaultseppunct}\relax
\EndOfBibitem
\bibitem[Fregoni \latin{et~al.}(2021)Fregoni, Haugland, Pipolo, Giovannini,
  Koch, and Corni]{fregoni2021strong}
Fregoni,~J.; Haugland,~T.~S.; Pipolo,~S.; Giovannini,~T.; Koch,~H.; Corni,~S.
  Strong coupling between localized surface plasmons and molecules by coupled
  cluster theory. \emph{Nano Letters} \textbf{2021}, \emph{21},
  6664--6670\relax
\mciteBstWouldAddEndPuncttrue
\mciteSetBstMidEndSepPunct{\mcitedefaultmidpunct}
{\mcitedefaultendpunct}{\mcitedefaultseppunct}\relax
\EndOfBibitem
\bibitem[Mondal \latin{et~al.}(2022)Mondal, Semenov, Ochoa, and
  Nitzan]{mondal2022strong}
Mondal,~M.; Semenov,~A.; Ochoa,~M.~A.; Nitzan,~A. Strong coupling in infrared
  plasmonic cavities. \emph{The Journal of Physical Chemistry Letters}
  \textbf{2022}, \emph{13}, 9673--9678\relax
\mciteBstWouldAddEndPuncttrue
\mciteSetBstMidEndSepPunct{\mcitedefaultmidpunct}
{\mcitedefaultendpunct}{\mcitedefaultseppunct}\relax
\EndOfBibitem
\bibitem[Jamshidi \latin{et~al.}(2023)Jamshidi, Kargar, Mendive-Tapia, and
  Vendrell]{jamshidi2023coupling}
Jamshidi,~Z.; Kargar,~K.; Mendive-Tapia,~D.; Vendrell,~O. Coupling molecular
  systems with plasmonic nanocavities: A quantum dynamics approach. \emph{The
  Journal of Physical Chemistry Letters} \textbf{2023}, \emph{14},
  11367--11375\relax
\mciteBstWouldAddEndPuncttrue
\mciteSetBstMidEndSepPunct{\mcitedefaultmidpunct}
{\mcitedefaultendpunct}{\mcitedefaultseppunct}\relax
\EndOfBibitem
\bibitem[Fregoni \latin{et~al.}(2022)Fregoni, Garcia-Vidal, and
  Feist]{fregoni2022theoretical}
Fregoni,~J.; Garcia-Vidal,~F.~J.; Feist,~J. Theoretical challenges in
  polaritonic chemistry. \emph{ACS photonics} \textbf{2022}, \emph{9},
  1096--1107\relax
\mciteBstWouldAddEndPuncttrue
\mciteSetBstMidEndSepPunct{\mcitedefaultmidpunct}
{\mcitedefaultendpunct}{\mcitedefaultseppunct}\relax
\EndOfBibitem
\bibitem[Wang and Dou(2023)Wang, and Dou]{wang2023nonadiabatic}
Wang,~Y.; Dou,~W. Nonadiabatic dynamics near metal surface with periodic
  drivings: A Floquet surface hopping algorithm. \emph{The Journal of Chemical
  Physics} \textbf{2023}, \emph{158}\relax
\mciteBstWouldAddEndPuncttrue
\mciteSetBstMidEndSepPunct{\mcitedefaultmidpunct}
{\mcitedefaultendpunct}{\mcitedefaultseppunct}\relax
\EndOfBibitem
\bibitem[Wang \latin{et~al.}(2024)Wang, Mosallanejad, Liu, and
  Dou]{wang2024nonadiabatic}
Wang,~Y.; Mosallanejad,~V.; Liu,~W.; Dou,~W. Nonadiabatic dynamics near metal
  surfaces with periodic drivings: A generalized surface hopping in Floquet
  representation. \emph{Journal of Chemical Theory and Computation}
  \textbf{2024}, \emph{20}, 644--650\relax
\mciteBstWouldAddEndPuncttrue
\mciteSetBstMidEndSepPunct{\mcitedefaultmidpunct}
{\mcitedefaultendpunct}{\mcitedefaultseppunct}\relax
\EndOfBibitem
\bibitem[Seidner \latin{et~al.}(1992)Seidner, Stock, Sobolewski, and
  Domcke]{seidner1992b}
Seidner,~L.; Stock,~G.; Sobolewski,~A.; Domcke,~W. A b initio characterization
  of the S 1--S 2 conical intersection in pyrazine and calculation of spectra.
  \emph{The Journal of chemical physics} \textbf{1992}, \emph{96},
  5298--5309\relax
\mciteBstWouldAddEndPuncttrue
\mciteSetBstMidEndSepPunct{\mcitedefaultmidpunct}
{\mcitedefaultendpunct}{\mcitedefaultseppunct}\relax
\EndOfBibitem
\bibitem[Woywod \latin{et~al.}(1994)Woywod, Domcke, Sobolewski, and
  Werner]{woywod1994characterization}
Woywod,~C.; Domcke,~W.; Sobolewski,~A.~L.; Werner,~H.-J. Characterization of
  the S 1--S 2 conical intersection in pyrazine using ab initio
  multiconfiguration self-consistent-field and multireference
  configuration-interaction methods. \emph{The Journal of chemical physics}
  \textbf{1994}, \emph{100}, 1400--1413\relax
\mciteBstWouldAddEndPuncttrue
\mciteSetBstMidEndSepPunct{\mcitedefaultmidpunct}
{\mcitedefaultendpunct}{\mcitedefaultseppunct}\relax
\EndOfBibitem
\bibitem[Kim \latin{et~al.}(2017)Kim, Hyla, Winget, Li, Wyss, Jordan, Larrain,
  Sadighi, Fuentes-Hernandez, Kippelen, \latin{et~al.}
  others]{kim2017reduction}
Kim,~H.~K.; Hyla,~A.~S.; Winget,~P.; Li,~H.; Wyss,~C.~M.; Jordan,~A.~J.;
  Larrain,~F.~A.; Sadighi,~J.~P.; Fuentes-Hernandez,~C.; Kippelen,~B.,
  \latin{et~al.}  Reduction of the work function of gold by N-heterocyclic
  carbenes. \emph{Chemistry of Materials} \textbf{2017}, \emph{29},
  3403--3411\relax
\mciteBstWouldAddEndPuncttrue
\mciteSetBstMidEndSepPunct{\mcitedefaultmidpunct}
{\mcitedefaultendpunct}{\mcitedefaultseppunct}\relax
\EndOfBibitem
\bibitem[Shi \latin{et~al.}(2010)Shi, Kim, Reina, Hofmann, Li, and
  Kong]{shi2010work}
Shi,~Y.; Kim,~K.~K.; Reina,~A.; Hofmann,~M.; Li,~L.-J.; Kong,~J. Work function
  engineering of graphene electrode via chemical doping. \emph{ACS nano}
  \textbf{2010}, \emph{4}, 2689--2694\relax
\mciteBstWouldAddEndPuncttrue
\mciteSetBstMidEndSepPunct{\mcitedefaultmidpunct}
{\mcitedefaultendpunct}{\mcitedefaultseppunct}\relax
\EndOfBibitem
\bibitem[Ma \latin{et~al.}(2023)Ma, Gong, Zeng, and Liu]{ma2023recent}
Ma,~Y.; Gong,~J.; Zeng,~P.; Liu,~M. Recent progress in interfacial dipole
  engineering for perovskite solar cells. \emph{Nano-Micro Letters}
  \textbf{2023}, \emph{15}, 173\relax
\mciteBstWouldAddEndPuncttrue
\mciteSetBstMidEndSepPunct{\mcitedefaultmidpunct}
{\mcitedefaultendpunct}{\mcitedefaultseppunct}\relax
\EndOfBibitem
\bibitem[Schneider \latin{et~al.}(1990)Schneider, Domcke, and
  K{\"o}ppel]{schneider1990aspects}
Schneider,~R.; Domcke,~W.; K{\"o}ppel,~H. Aspects of dissipative electronic and
  vibrational dynamics of strongly vibronically coupled systems. \emph{The
  Journal of chemical physics} \textbf{1990}, \emph{92}, 1045--1061\relax
\mciteBstWouldAddEndPuncttrue
\mciteSetBstMidEndSepPunct{\mcitedefaultmidpunct}
{\mcitedefaultendpunct}{\mcitedefaultseppunct}\relax
\EndOfBibitem
\bibitem[Shirley(1965)]{shirley1965solution}
Shirley,~J.~H. Solution of the Schr{\"o}dinger equation with a Hamiltonian
  periodic in time. \emph{Physical Review} \textbf{1965}, \emph{138},
  B979\relax
\mciteBstWouldAddEndPuncttrue
\mciteSetBstMidEndSepPunct{\mcitedefaultmidpunct}
{\mcitedefaultendpunct}{\mcitedefaultseppunct}\relax
\EndOfBibitem
\bibitem[Mosallanejad \latin{et~al.}(2023)Mosallanejad, Chen, and
  Dou]{mosallanejad2023floquet}
Mosallanejad,~V.; Chen,~J.; Dou,~W. Floquet-driven frictional effects.
  \emph{Physical Review B} \textbf{2023}, \emph{107}, 184314\relax
\mciteBstWouldAddEndPuncttrue
\mciteSetBstMidEndSepPunct{\mcitedefaultmidpunct}
{\mcitedefaultendpunct}{\mcitedefaultseppunct}\relax
\EndOfBibitem
\bibitem[Leskes \latin{et~al.}(2010)Leskes, Madhu, and Vega]{leskes2010floquet}
Leskes,~M.; Madhu,~P.; Vega,~S. Floquet theory in solid-state nuclear magnetic
  resonance. \emph{Progress in nuclear magnetic resonance spectroscopy}
  \textbf{2010}, \emph{57}, 345--380\relax
\mciteBstWouldAddEndPuncttrue
\mciteSetBstMidEndSepPunct{\mcitedefaultmidpunct}
{\mcitedefaultendpunct}{\mcitedefaultseppunct}\relax
\EndOfBibitem
\bibitem[Dou and Subotnik(2017)Dou, and Subotnik]{dou2017generalized}
Dou,~W.; Subotnik,~J.~E. A generalized surface hopping algorithm to model
  nonadiabatic dynamics near metal surfaces: The case of multiple electronic
  orbitals. \emph{Journal of chemical theory and computation} \textbf{2017},
  \emph{13}, 2430--2439\relax
\mciteBstWouldAddEndPuncttrue
\mciteSetBstMidEndSepPunct{\mcitedefaultmidpunct}
{\mcitedefaultendpunct}{\mcitedefaultseppunct}\relax
\EndOfBibitem
\bibitem[Nenner and Schulz(1975)Nenner, and Schulz]{nenner1975temporary}
Nenner,~I.; Schulz,~G. Temporary negative ions and electron affinities of
  benzene and N-heterocyclic molecules: pyridine, pyridazine, pyrimidine,
  pyrazine, and s-triazine. \emph{The Journal of Chemical Physics}
  \textbf{1975}, \emph{62}, 1747--1758\relax
\mciteBstWouldAddEndPuncttrue
\mciteSetBstMidEndSepPunct{\mcitedefaultmidpunct}
{\mcitedefaultendpunct}{\mcitedefaultseppunct}\relax
\EndOfBibitem
\bibitem[Song \latin{et~al.}(2002)Song, Lee, and Kim]{song2002photoelectron}
Song,~J.~K.; Lee,~N.~K.; Kim,~S.~K. Photoelectron spectroscopy of pyrazine
  anion clusters. \emph{The Journal of chemical physics} \textbf{2002},
  \emph{117}, 1589--1594\relax
\mciteBstWouldAddEndPuncttrue
\mciteSetBstMidEndSepPunct{\mcitedefaultmidpunct}
{\mcitedefaultendpunct}{\mcitedefaultseppunct}\relax
\EndOfBibitem
\bibitem[Palihawadana \latin{et~al.}(2012)Palihawadana, Sullivan, Buckman, and
  Brunger]{palihawadana2012electron}
Palihawadana,~P.; Sullivan,~J.; Buckman,~S.; Brunger,~M. Electron scattering
  from pyrazine: Elastic differential and integral cross sections. \emph{The
  Journal of Chemical Physics} \textbf{2012}, \emph{137}\relax
\mciteBstWouldAddEndPuncttrue
\mciteSetBstMidEndSepPunct{\mcitedefaultmidpunct}
{\mcitedefaultendpunct}{\mcitedefaultseppunct}\relax
\EndOfBibitem
\bibitem[Arenas \latin{et~al.}(1996)Arenas, Woolley, Otero, and
  Marcos]{arenas1996charge}
Arenas,~J.~F.; Woolley,~M.~S.; Otero,~J.~C.; Marcos,~J.~I. Charge-transfer
  processes in surface-enhanced Raman scattering. Franck- condon active
  vibrations of pyrazine. \emph{The Journal of Physical Chemistry}
  \textbf{1996}, \emph{100}, 3199--3206\relax
\mciteBstWouldAddEndPuncttrue
\mciteSetBstMidEndSepPunct{\mcitedefaultmidpunct}
{\mcitedefaultendpunct}{\mcitedefaultseppunct}\relax
\EndOfBibitem
\bibitem[Gu and Mukamel(2020)Gu, and Mukamel]{gu2020manipulating}
Gu,~B.; Mukamel,~S. Manipulating nonadiabatic conical intersection dynamics by
  optical cavities. \emph{Chemical science} \textbf{2020}, \emph{11},
  1290--1298\relax
\mciteBstWouldAddEndPuncttrue
\mciteSetBstMidEndSepPunct{\mcitedefaultmidpunct}
{\mcitedefaultendpunct}{\mcitedefaultseppunct}\relax
\EndOfBibitem
\end{mcitethebibliography}

\begin{center}
{\bfseries \Large Supporting Information: }

{\bfseries \Large Manipulating nonadiabatic dynamics by plasmonic nanocavity}
\end{center}

\begin{figure}[htbp]
    \centering
    \includegraphics[width=0.8\linewidth]{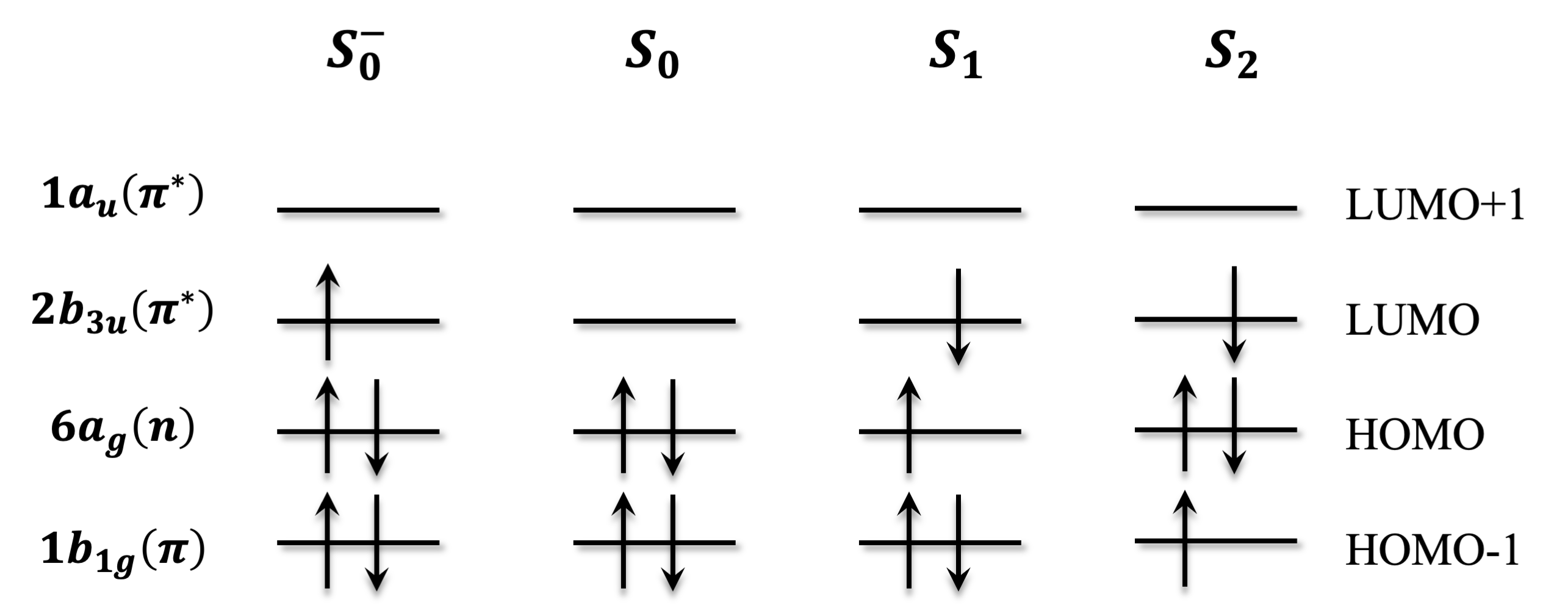}
    \caption{Electronic configurations of $S_0^-$, $S_0$, $S_1$ and $S_2$ states. Here, HOMO means the highest occupied molecular orbital, and LUMO means the lowest unoccupied molecular orbital.} 
    \label{fig:pes}
\end{figure}

\begin{figure}[htbp]
    \centering
    \includegraphics[width=0.8\linewidth]{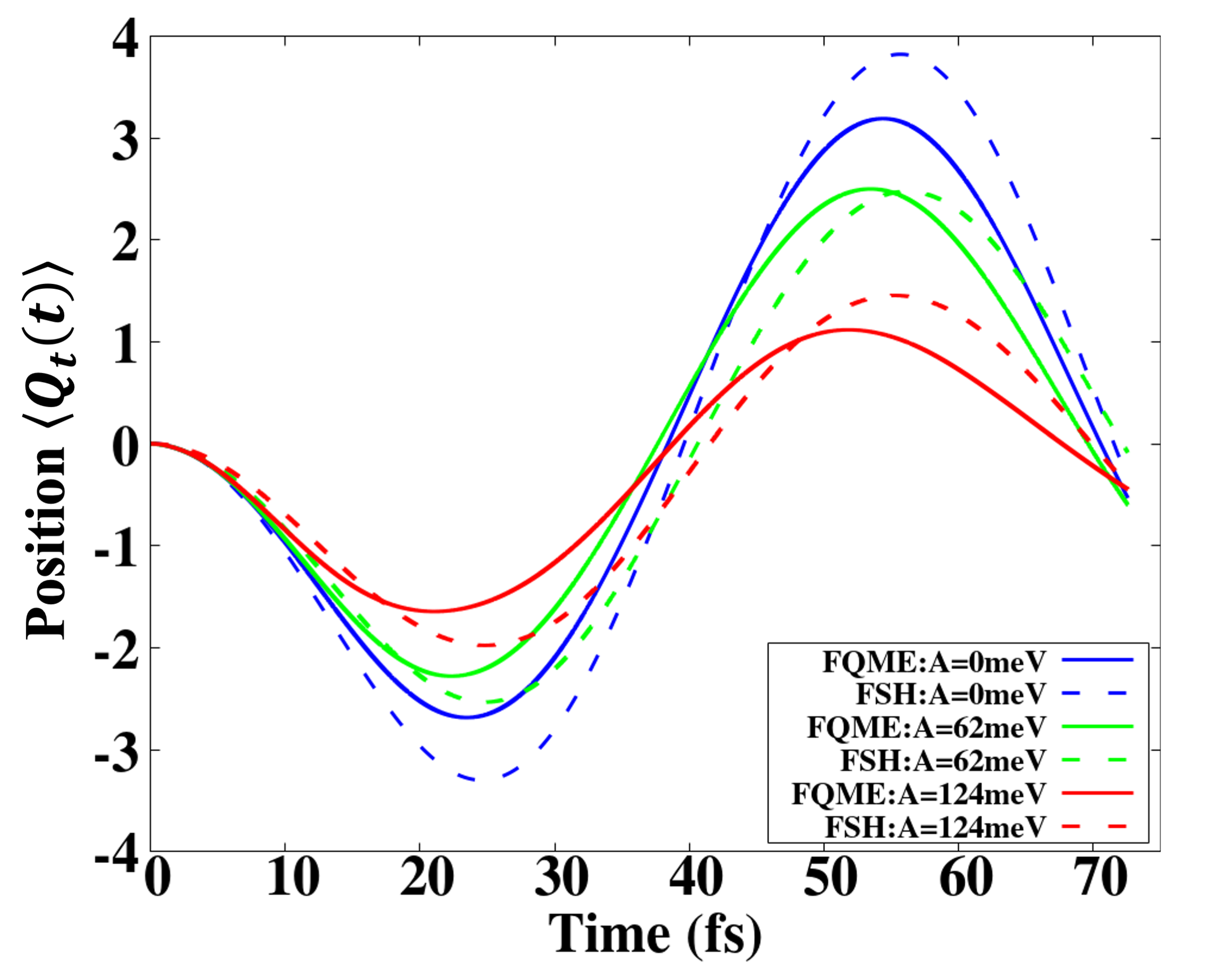}
    \caption{Nuclear dynamics along the tuning mode under different driving conditions.}
    \label{fig:pes}
\end{figure}

\begin{figure}[htbp]
    \centering
    \includegraphics[width=1.0\linewidth]{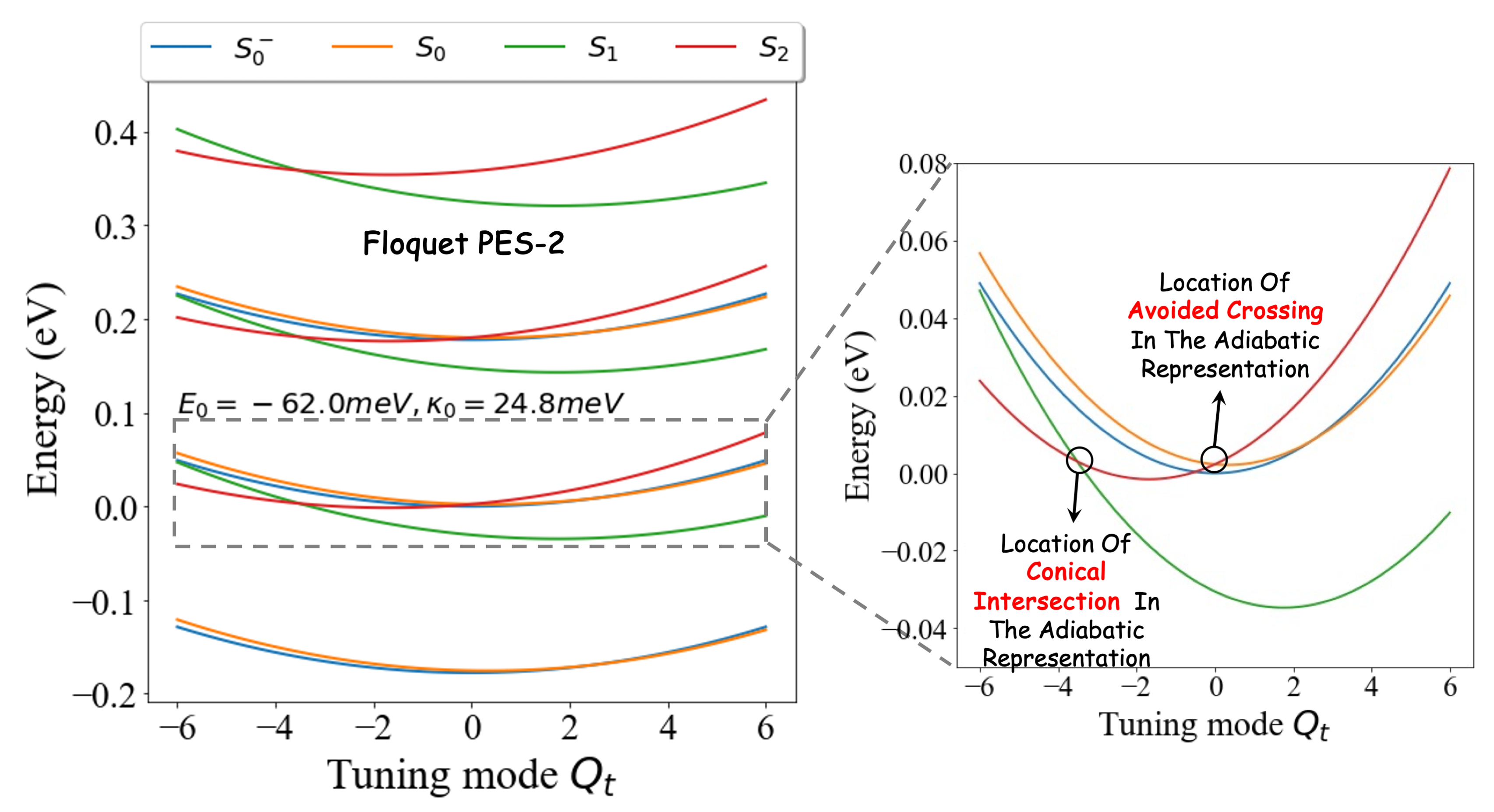}
    \caption{The Floquet diabatic potential energy surfaces of $S_0^-, S_0, S_1, S_2$ states along the tuning mode. Here, we consider three Floquet levels, which means each state has three Floquet replicas with energy difference $\hbar\Omega$. The inset highlights the location of conical intersection in the adiabatic representation due to the coupling between $S_2$ and $S_1$ states, as well as the location of avoided crossing in the adiabatic representation due to the coupling between $S_2$ and $S_0$ states.}
    \label{fig:pes}
\end{figure}







\end{document}